\documentstyle[11pt]{article}
       \font\tenmsb=msbm10
       \font\sevenmsb=msbm7
       \font\fivemsb=msbm5     
       \catcode`\@=11
       \ifx\amstexloaded@\relax\catcode`\@=\active
       \endinput\else\let\amstexloaded@\relax\fi
       \def\spaces@{\space\space\space\space\space}
       \def\spaces@@{\spaces@\spaces@\spaces@\spaces@\spaces@}
       \def\space@.{\futurelet\space@\relax}
       \space@. %
       \def\Err@#1{\errhelp\defaulthelp@\errmessage{AmS-TeX error: #1}}
       \def\relaxnext@{\let\next\relax}
       \def\accentfam@{7}
       \def\noaccents@{\def\accentfam@{0}}
       \def\Cal{\relaxnext@\ifmmode\let\next\Cal@\else
       \def\next{\Err@{Use \string\Cal\space only in math mode}}\fi\next}
       \def\Cal@#1{{\Cal@@{#1}}}
       \def\Cal@@#1{\noaccents@\fam\tw@#1}
       \def\Bbb{\relaxnext@\ifmmode\let\next\Bbb@\else
       \def\next{\Err@{Use \string\Bbb\space only in math mode}}\fi\next}
       \def\Bbb@#1{{\Bbb@@{#1}}}
       \def\Bbb@@#1{\noaccents@\fam\msbfam#1}
       \newfam\msbfam
       \textfont\msbfam=\tenmsb
       \scriptfont\msbfam=\sevenmsb
       \scriptscriptfont\msbfam=\fivemsb

\def\N{{\Bbb N}}
\def\Z{{\Bbb Z}}
\def\R{{\Bbb R}}
\def\T{{\Bbb T}}
\def\C{{\Bbb C}}

\def\skipaline{\removelastskip\vskip12pt plus 1pt minus 1pt}

\def\Proof{\removelastskip\skipaline
\noindent \it Proof. \rm}

\newtheorem{Theorem}{Theorem}
\newtheorem{Lemma}{Lemma}[section]
\newtheorem{Proposition}{Proposition}[section]

\@addtoreset{equation}{section}

\newcommand{\sss}{\smallskip\noindent}
\newcommand{\ms}{\medskip\noindent}
\newcommand{\bs}{\bigskip\noindent}

\newcommand{\qed}{\nolinebreak\hfill\rule{2mm}{2mm}
\par\medbreak}

\newcommand{\lep}{ \;<\;{\rm c}\; }
\newcommand{\la}{\langle }
\newcommand{\ra}{\rangle }
\newcommand{\ko}{\langle k, \omega\rangle }
\newcommand{\gep}{\;>{\rm c}\; }
\newcommand{\kth}{e^{{\rm i}\langle k, \theta\rangle}\; }

\newcommand{\nx}{e^{{\rm i}\langle n, x\rangle}\; }
\newcommand{\beq}{\begin{equation} }
\newcommand{\eeq}{\end{equation} }

\newcommand{\gammaa}{\gamma^{-{\rm c}}}

\newcommand{\kk}{|k|^{\rm c}}
\newcommand{\gammaaa}{\gamma^{-{\rm c}}}
\newcommand{\gammab}{-\rm c}

\begin{document} 
\setlength{\columnsep}{5pt}

\title{ KAM Tori for 1D 
Nonlinear Wave Equations with Periodic Boundary Conditions}
\author{\\ Luigi Chierchia
\\{\footnotesize Dipartimento di Matematica}
\\{\footnotesize Universit\`a Roma Tre}
\\{\footnotesize E-mail: luigi@matrm3.mat.uniroma3.it} 
\\
\\  Jiangong You\thanks{The research was partly supported by NNSF of China and by
the italian CNR grant \# 211.01.31.}
\\{\footnotesize Department of Mathematics}
\\{\footnotesize Nanjing University, Nanjing 210093,  China}
\\{\footnotesize E-mail: jyou@ nju.edu.cn}}  
\maketitle

\date{October, 1998}
\maketitle

\begin{abstract}

\noindent
In this paper, one--dimensional (1D)  nonlinear wave equations 
$$
u_{tt}
-u_{xx}+V(x)u
=f(u),
$$
with  periodic boundary conditions are considered; $V$ is a periodic smooth
or analytic function and the nonlinearity $f$ is an analytic function
vanishing together with its derivative at $u=0$. It is proved that for 
``most''  potentials $V(x)$, the above equation admits
small-amplitude periodic or quasi-periodic solutions corresponding to finite 
dimensional invariant tori for an associated infinite dimensional dynamical system.
The proof is based  on an infinite dimensional KAM theorem which 
allows for multiple normal frequencies.

\end{abstract}

\section{Introduction and Results}

In the 90's, the celebrated KAM (Kolmogorov-Arnold-Moser) theory has been 
successfully extended to infinite dimensional settings  so as to deal with certain 
classes of partial differential equations carrying a Hamiltonian structure,
including, as a typical example,  wave equations of the form 
\begin{equation}\label{1.1}
 u_{tt}-u_{xx}+ V(x)u= f(u),\quad f(u)=O(u^2),\eeq
see Wayne \cite{Wa}, Kuksin \cite{K} and P\"oschel \cite{P1}. In such 
papers, KAM theory for lower dimensional tori \cite{M1}, \cite{Me}, \cite{E} 
(i.e., invariant tori of dimension lower than the number of degrees of freedom),
 has been generalized in order to prove
the existence of small-amplitude quasi-periodic solutions for 
(\ref{1.1}) subject to Dirichlet or Neumann boundary conditions (on a 
finite interval for odd and analytic nonlinearities $f$). The technically 
more difficult periodic boundary condition case has been later considered by
Craig and Wayne \cite{CW} who established the existence of periodic solutions.
 The techniques used in \cite{CW} are based not on KAM theory, but rather
 on a generalization of the Lyapunov-Schmidt procedure and on  techniques by
Fr\"ohlich and Spencer \cite{FS}. Recently, Craig and Wayne's approach
has been significantly improved by Bourgain \cite{B1}, \cite{B2}
who obtained  the existence of quasi-periodic solutions for certain kind
of 1D and, most notably,  2D partial differential equations with periodic boundary 
conditions.

\sss
The technical reason why KAM theory has not been used to treat the periodic 
boundary condition case is related to the multiplicity of the spectrum of the 
Sturm-Liouville  operator $A=-\frac{d^2}{dx^2}+V(x)$. Such multiplicity leads
to some extra ``small denominator"problems related to the so called normal
frequencies.

\sss
The purpose of this paper is to show that, allowing for more general normal
forms,  one  can indeed use KAM techniques  to deal also with the multiple 
normal frequency case arising in PDE's with periodic boundary conditions. 

\sss
A rough description of our results is as follows.
Consider the periodic boundary problem for (\ref{1.1})
with an analytic nonlinearity $f$ and a real analytic (or smooth enough)
potential $V$. Such potential will be taken in a
$d$-dimensional family of functions parameterized by a real $d$--vector $\xi$,
$V(x)=V(x,\xi)$, satisfying 
general non-degenerate (``non--\-resonance--\-of--\-eigenvalue")
conditions. Then for ``most" potentials in the family (i.e. for most $\xi$ in
Lebesgue measure sense), there exist small-amplitude
quasi-periodic solutions for (\ref{1.1}) corresponding to 
$d$-dimensional KAM tori   for the associated infinite dimensional 
Hamiltonian system. Moreover (as usual in the  KAM approach) 
one obtains, for the constructed solutions, a
local normal form  which provides  linear
stability in the case the  operator $A$ is positive definite.

\sss
We believe that the technique used in this paper can be generalized so 
as to cover 2D wave equations.

\sss
The paper is organized as follows: In section 2 we formulate a general 
infinite dimensional KAM Theorem designed to deal with multiple normal
frequency cases; in section 3 we show how to apply the preceding KAM Theorem
to the nonlinear wave equation (\ref{1.1}) with periodic boundary 
conditions. The proof of  the KAM Theorem is provided in sections 4$\div$6. Some 
technical lemmata are proved in the Appendix.

\section{An infinite dimensional KAM Theorem}
In this section we will formulate a KAM Theorem in an infinite dimensional
setting which can be applied to some 1D  partial differential equations  with
periodic boundary conditions.

\sss
We start by introducing some notations.

\subsection{ Spaces}

For $n\in \Bbb N$, let  $d_n\in \Bbb Z_+$ be positive {\sl even}
integers\footnote{We use the notations 
$\Bbb N=\{0, 1,2,\cdots\}, \Bbb Z_+=\{1,2,\cdots\}$.}.
Let $\Cal Z\equiv \otimes_{n\in \Bbb N} \C^{d_n}$: the coordinates in 
$\Cal Z$ are given by 
$z=(z_0,z_1, z_2, \cdots)$  with
$z_n\equiv (z_n^1, \cdots, z_n^{d_n})\in \C^{d_n}$.
 Given two real numbers $a, \rho$, we consider the
 (Banach) subspace of $\Cal Z$ given by
 \[\Cal Z_{a,\rho}=\{z\in \Cal Z: \ \ |z|_{a,\rho} <\infty \}\]
where the norm $|\cdot |_{a,\rho}$ is defined as
\[ |z|_{a,\rho}= |z_0|+\sum_{n\in \Z_+} |z_n|n^a e^{n\rho},\]
(and the norm in $\C^{d_n}$ is taken to be the 1--norm  $|z_n|=
\sum_{j=1}^{d_n}|z_n^j|$).

\sss
{\sl In what follows, we shall consider  either $a= 0$ and $\rho >0$ or $a>0$ and
$ \rho=0$
 (corresponding respectively to the analytic case or the finitely smooth case)}.
   
\sss
The role of complex neighborhoods in phase space of KAM theory will be 
played here  by the  set
\[ \Cal P_{a, \rho}\equiv \hat {\Bbb T}^d\times \Bbb C^d\times 
\Cal Z_{a, \rho},\]
where $\hat{\Bbb T}^d$ is the complexification of the real torus
$\Bbb T^d= \Bbb R^d/2\pi\Bbb Z^d$. 

\sss
For positive numbers $r, s$ we denote by
\begin{equation}
D_{a, \rho}(r, s)=\{(\theta,I,z)\in \Cal P_{a,\rho}: 
 |{\rm Im}\, \theta|<r, |I|<s^2,
 |z|_{a,\rho}<s\}
\end{equation}
a complex neighborhood of $\Bbb T^d\times\{I=0\}\times\{ z=0\}$.
Finally, we denote by  $\Cal O$  a given compact set in $\Bbb R^d$
with positive Lebesgue measure: $\xi\in \Cal O$ will parameterize a selected
family of potential $V=V(x, \xi)$ in (\ref{1.1}).

\subsection { Functions }

We consider functions $F$ on $D_{a,\rho}(r,s)\times \Cal O$ having the following
properties: (i) $F$ is real for real arguments; (ii) $F$ admits an expansion of the
form
\beq\label{Fa}
F=\sum_\alpha F_\alpha z^\alpha,\eeq
where the multi-index 
$\alpha$ runs over the set
$\alpha\equiv (\alpha_0,\alpha_1,...)
\in \otimes_{n\in\Bbb N}\Bbb N^{d_n}$ with finitely many
non-vanishing  components\footnote{Thus $\exists$ $n_0>0$ such that
$\displaystyle z^\alpha\equiv \prod_{n=0}^{n_0} z_n^{\alpha_n}\equiv 
\prod_{n=0}^{n_0} \prod_{j=1}^{d_n} (z_n^j)^{\alpha_n^j}$.}
$\alpha_n$; (iii) for each $\alpha$, the
function
$F_\alpha=F_\alpha(\theta, I, \xi)$ is real analytic in the variables
$(\theta, I)\in
\{ |{\rm Im} \theta|<r, |I|<s^2\}$; (iv) for each $\alpha$, the dependence of
$F_\alpha$ upon the parameter $\xi$ is of class $C^{\bar d}_W(\Cal O)$ for
some $\bar d>0$ (to be fixed later): here $C^m_W(\Cal O)$ denotes the class of
functions which are
$m$ times differentiable on the closed set $\cal O$ in the sense of Whitney
\cite{Whit}.

\sss
The convergence of the expansion (\ref{Fa}) in $D_{a,\rho}(r,s)\times \Cal O$
will be
guaranteed by assuming the finiteness of the following weighted norm: 
\begin{equation}\label{Fnorm}
  \|F\|_{D_{a,\rho}(r,s),\Cal O}\equiv\sup_{|z|_{a,\rho}\le s}\sum_{\alpha}
\|F_\alpha\|\,\, |z^\alpha|
\end{equation}
where, if $\displaystyle F_\alpha=\sum_{k\in \Z^d, l\in \N^d} F_{kl\alpha}(\xi)I^l
\kth$, $\|F_\alpha\|$ is defined as
\beq\label{Falpha}\|F_\alpha\|\equiv
 \sum_{k, l}
|F_{kl\alpha}|_{\Cal O}\,
s^{2|l|} e^{|k|r},\qquad
|F_{kl\alpha}|_{\Cal O}\equiv
\max_{|p|\le \bar d^2}|\frac{\partial^pF_{kl\alpha}}
{\partial\xi^p}|,\eeq
(the derivatives with respect to $\xi$ are in the sense of Whitney).

\sss
The  set of functions $F: D_{a,\rho}(r,s) \times \Cal O\to \C$ verifying
(i)$\div$(iv) above  with finite
$\|\cdot\|_{ D_{a,\rho}(r,s),\cal O}$ norm
will be denoted by $\Cal F_{D_{a, \rho}(r,s), \Cal O}$.

\subsection {Hamiltonian vector fields and Hamiltonian equations}

To functions $F\in \Cal F_{D_{a, \rho}(r,s),\cal O}$, we associate a Hamiltonian 
vector field defined as
\[X_F =(F_I, -F_\theta, \{{\rm i}J_{d_n}F_{z_n}\}_{n\in \Bbb N}),\]
where $J_{d_n}$ denotes the standard symplectic matrix 
 $
\left(\begin{array}{cc}
0 & I_{d_n/2}\\
-I_{d_n/2} & 0
\end{array} \right )$
and  ${\rm i}=\sqrt{-1}$; the derivatives of $F$ are defined as the derivatives
term--by--term of the series (\ref{Fa}) defining $F$. The appearence of the imaginary
unit is due to notational convenience and will be justified later by the use of
complex canonical variables.

\sss 
Correspondingly we consider the
Hamiltonian equations\footnote{Dot stands for
the time derivatives $d/dt$.}
\begin{equation}\label{2.5}
\dot \theta= F_I, \quad \dot I=-F_\theta,\quad
\dot z_n
= {\rm i}J_{d_n}F_{z_n},\quad n\in \Bbb N.
\end{equation}
{\sl A solution of such equation is intended to be just a $C^1$ map from an interval
to the domain of definition of $F$, $D_{a,\rho}(r,s)$, satisfying} (\ref{2.5}).

\sss
Given a real number $\bar a$, we define also a  weighted norm for $X_ F$ by 
letting\footnote{The norm  $\|\cdot\|_{D_{a, \rho}( r,s), \cal O}$ for scalar
functions is defined in (\ref{Fnorm}). For vector (or matrix--valued) functions $G:
D_{a, \rho}( r,s)\times {\cal O}\to \C^m$, ($m<\infty$) is similarly defined as
$\|G\|_{D_{a, \rho}( r,s), \cal O}=\sum_{i=1}^m\|G_i\|_{D_{a, \rho}( r,s), \cal O}$ 
(for the matrix--valued case the sum will run over all entries).}
\begin{eqnarray}\label{weightednorm}
&&\|X_F\|_{\!{}_{D_{a, \rho}(r,s), \cal O}}^{\bar a,\rho}\equiv  \\
\quad &&\|F_I\|_{\!{}_{D_{a, \rho}(r,s), \cal O}}+
\frac 1{s^2}\|F_\theta\|_{\!{}_{D_{a, \rho}(r,s), \cal O}}+
\frac 1s(\|F_{z_0}\|_{\!{}_{D_{a, \rho}(r,s), \cal O}}+
\sum_{n\in \Z_+} \|F_{z_n}\|_{\!{}_{D_{a, \rho}(r,s), \cal O}}n^{\bar a} e^{n\rho}).
\nonumber\end{eqnarray}

\noindent{\bf Notational Remark} In what follows,
 only the indices $r, s$ and the set $\Cal O$ will change while $a, \bar a, \rho$
will be kept fixed, therefore we shall usually denote 
$\|X_F\|_{D_{a,\rho}(r,s),\Cal O}^{\bar a,\rho}$ by $
\|X_F\|_{r,s,\Cal O}$, $D_{a, \rho}(r,s)$ by $D(r,s)$ and
$\Cal F_{D_{a, \rho}(r,s), \Cal O}$ by $\Cal F_{r,s,\Cal O}$.

\ms
\subsection{Perturbed Hamiltonians and the KAM result}

The starting point will be  a family of  integrable Hamiltonians 
of the form 
\begin{equation}\label{hamN} 
N =\la\omega(\xi),I\ra+ \frac 12\sum_{n\in \N}\la A_n(\xi)z_n, z_n\ra,
\end{equation}
where  $\xi\in \Cal O$ is a parameter, $A_n$ is a $d_n\times
d_n$ real symmetric matrix and $\la\cdot,\cdot\ra$ is the standard inner product;
here the phase space ${\cal P}_{a,\rho}$ is endowed with the symplectic form
$\displaystyle dI\wedge d\theta + {\rm i} \sum_n \sum_{j=1}^{d_n/2} z_n^j \wedge d
z_n^{j+d_n/2}$. 

\sss
{\sl For simplicity, we shall take, later,  $\omega(\xi)\equiv \xi$}.

\sss
For each $\xi\in \Cal O$, the Hamiltonian equations of motion for $N$, i.e.,  
\begin{equation}\label{unperturbed}
\frac {d\theta}{dt}= \omega,\quad  \frac {dI}{dt}=0, \quad\frac {dz_n}{dt}
= {\rm i}J_{d_n}A_n  z_n, \quad n\in \Bbb N,
\end{equation} 
admit special solutions 
$(\theta, 0, 0)\to (\theta+\omega t, 0, 0)$  corresponding to 
 an invariant torus in $\Cal P_{a, \rho}$. 

\sss
Consider now  the perturbed Hamiltonians
\begin{equation}\label{hamH}
H=N+P =\la\omega(\xi),I\ra+ \frac 12
\sum_{n\in \Bbb N}\la A_n(\xi)z_n, z_n\ra+ P(\theta,I,z, \xi)
\end{equation} with  $P\in \Cal F_{r,s,\Cal O}$.

\sss
Our goal is to prove that, for most values of parameter
 $\xi \in \Cal O$ (in Lebesgue measure 
sense), the Hamiltonian $H=N+P$ still admits 
 an invariant torus provided  $\|X_P\|$ is sufficiently
small. 

\ms
In order to obtain this kind of result {\sl we 
make the following assumptions on $A_n$ and the perturbation $P$}.

\bs
\noindent
(A1) {\it Asymptotics of  eigenvalues:}
There exist $\bar d\in \Bbb N, \delta>0$ and $b\ge 1$ such that
$d_n \le \bar d $ for all $n$, and 
\begin{equation}\label{asymp1}
A_n= \lambda_n 
\left(\begin{array}{cc}
0 & I_{d_n/2}\\
I_{d_n/2} & 0
\end{array}\right)
 + B_n,\quad B_n=O(n^{-\delta})
\end{equation}
where $\lambda_n$ are real and independent of $\xi$ while $B_n$ may
depend on $\xi$; furthermore, the behaviour of $\lambda_n$'s is assumed 
to be as follows
\begin{equation} \label{asymp2}
\lambda_n= n^b+o(n^b), \quad\frac {\lambda_m-\lambda_n}{m^b-n^b}
= 1+ o(n^{-\delta}), \ \ \ n< m.\end{equation}

 \bs
\noindent
(A2) {\it Gap condition:}\ There exists $\delta_1>0$ such that 
\[{\rm dist}\left(\sigma(J_{d_i}A_{i}),
\sigma(J_{d_j}A_j)\right)> \delta_1>0,\quad \forall i\ne j\ ;\] 
($\sigma(\cdot)$ denotes ``spectrum of $\cdot$").

\ms
Note that, for large $i,j$, the {\it gap condition} follows
from the asymptotic property. 

\bs
\noindent
(A3) {\it Smooth dependence on parameters:}
All entries of $B_n$ are $\bar d^2$ Whitney--smooth functions of $\xi$ with 
 $C^{\bar d^2}_W$-norm  bounded  by some positive constant $L$.

\bs
\noindent
(A4) {\it Non-resonance condition:}
\begin{equation}\label{ndc}
   {\rm meas}\{\xi\in \Cal O:\quad
\la k, \omega(\xi)\ra(
\la k, \omega(\xi)\ra+\lambda(\xi))(\la
k,\omega(\xi)\ra+\lambda(\xi)+\mu(\xi))=0\}=0,
\end{equation}
for each $0\ne k\in \Bbb Z^d$ 
and for any  $  \lambda, \mu\in \bigcup_{n\in \Bbb N}\sigma(J_{d_n}A_n)$; meas
$\equiv$ Lebesgue measure.

\bs
\noindent
(A5) {\it Regularity of the perturbation}:
The perturbation $P\in \Cal F_{D_{a, \rho}(r,s), \Cal O}$ is {\sl regular} in the
sense that $\|X_P\|_{\!{}_{D_{a, \rho}(r,s), \cal O}}^{\bar a,\rho}<\infty$ 
with $\bar a>a$. In fact, we assume that one of the following holds:
\[ (a)\;\;\; \rho>0,\quad \bar a>a=0; \quad (b)\;\;\; \rho=0,\quad \bar a>a>0,\]
(such conditions correspond, respectively, to analytic or smooth  
solutions). In the case of $d=1$ (i.e., the periodic solution case) one can 
allow $\bar a=a$.

\bs
Now we can  state our KAM result.

\begin{Theorem}\label{KAM}
Assume that $N$ in (\ref{hamN}) satisfies (A1) - (A4) and $P$ is regular
in the  sense of (A5) and let $\gamma>0$.
There exists a positive constant 
$\epsilon=\epsilon(d, \bar d, b,\delta, \delta_1,\bar a-a, L, \gamma)$ 
such that if $\|X_P\|_{\!{}_{D_{a, \rho}(r,s), \cal O}}^{\bar
a,\rho}<\epsilon$, then the following holds true.
There exists a Cantor set 
$\Cal O_\gamma\subset\Cal O$ with ${\rm meas}(\Cal O\setminus \Cal O_\gamma)
\to 0$ as $\gamma\to 0$, and two maps (real analytic in $\theta$ and Whitney 
smooth in $\xi\in \Cal O$) 
$$\Psi: T^d\times \Cal O_\gamma\to D_{a,\rho}(r,s)\subset \Cal P_{a,\rho},\ \ \ \
\tilde \omega:\Cal O_\gamma\to R^d,$$
such that for any $\xi\in \Cal O_\gamma$ and $\theta\in T^d$
the curve $t\to \Psi(\theta+\tilde \omega(\xi) t,\xi)$ is a quasi-periodic 
solution of the Hamiltonian equations governed by $H=N+P$. Furthermore, 
$\Psi(\T^d, \xi)$ is a smoothly embedded $d$-dimensional $H$-invariant
torus in $\Cal P_{a,\rho}$.
\end{Theorem}

\ms
\noindent
{\bf Remarks} 
(i) For simplicity we shall in fact assume that {\sl all eigenvalues $\lambda_i$ of
$A_n$ are positive for all $n$'s}. The case of some non positive eigenvalues can be
easily dealt with at the expense of a (even) heavier notation.

\ms
(ii)
In the above case (i.e. positive eigenvalues), Theorem 1 yields  {\sl linearly
stable} KAM tori.

\ms 
\noindent
(iii)  The parameter $\gamma$  plays the role of the Diophantine constant for
the frequency $\tilde \omega$ in the sense that,
there is $\tau>0$ such that $\forall k\in \Z^d\backslash \{0\},$
 \[\la k, \tilde \omega\ra
>\frac {\gamma}{2|k|^\tau}.\]
Notice also that $\Cal O_\gamma$ is claimed to be nonempty and big only for 
$\gamma$ small enough.

\ms
\noindent
(iv)
The regularity property $\bar a>a$ is needed for the measure estimates on $\Cal
O\backslash \Cal O_\gamma$.
As we already  mentioned, such regularity requirement
is not necessary for periodic solution case, i.e., $d=1$. Thus 
{\sl the above theorem applies immediately to the  construction of  periodic
solutions for nonlinear Schr\"odinger equations}.

\ms
\noindent
(v). The non-degeneracy condition (\ref{ndc}) (which
is stronger than Bourgain's non-degenerate condition \cite{B2} but  weaker than
Melnikov's one \cite{Me}) covers the multiple normal frequency case: this is the
technical reason that allows to treat PDE's with periodic boundary conditions.

\section{Application to 1D wave equations}

\noindent
In this section we show how Theorem \ref{KAM} implies the existence
of quasi-periodic solutions for 1D wave
equations with periodic boundary conditions.

\sss
Let us rewrite the wave equation (\ref{1.1}) as follows
\begin{eqnarray}\label{nonlinearwave}
& & u_{tt}+Au= f(u),\ \ \
Au\equiv
-u_{xx}+V(x,\xi)u, \ x, t\in \R, \nonumber\\
& & u(t,x)=u(t, x+2\pi),\quad u_t(t,x)=u_t(t, x+2\pi),
\end{eqnarray}
where $V(\cdot,\xi)$ is a {\it real--analytic}
({\it  or smooth}) periodic potential
parameterized by some $\xi\in \R^d$ (see below) and
$f(u)$  is a  {\it real--analytic}
function near $u=0$ with $f(0)=f'(0)=0$.

\sss
As it is well known, the operator   $A$ 
with periodic boundary conditions admits an orthonormal basis of 
eigenfunctions $\phi_n\in L^2(\T), n\in \N$, with corresponding eigenvalues 
$\mu_n$ satisfying the following asymptotics for large $n$
\[\mu_{2n-1}, \mu_{2n}= n^2+\frac 1{2\pi}
\int_\T V(x)dx+O(n^{-2}).\]

\sss
{\sl For simplicity, we shall consider the case 
of vanishing mean value of the potential $V$ and assume that all eigenvalues are
positive:}
\begin{equation}\label{assum}
\int_\T V(x)dx=0\ ,\qquad
\mu_n\equiv \lambda_n^2>0\ ,\quad\forall\ n.
\end{equation}

\sss
Following Kuksin \cite{K} and Bourgain \cite{B1}, we consider a family of 
real analytic (or smooth) potentials
$V(x, \xi),$ where {\sl the $d$-parameters
$\xi=(\xi_1, \cdots, \xi_d)\in \Cal O\subset \R^d$ are simply taken 
to be a given set of $d$ frequencies $\lambda_{n_i}\equiv\sqrt{\mu_{n_i}}$}:
\begin{equation}\label{xml}
\xi_i\equiv\sqrt{\mu_{n_i}}\equiv\lambda_{n_i}\ , \qquad i=1, 
\cdots, d
\end{equation}
where $\mu_{n_i}$ are (positive) eigenvalues of\footnote{Plenty of such potentials
may be constructed with, e.g., the inverse spectral theory.} 
$A$. 

\sss
We may also (and shall) require that there exists a positive $\delta_1>0$ such that
\begin{equation}\label{del1}
|\mu_k-\mu_h|> \delta_1\ ,
\end{equation}
for all $k>h$ except when $k$ is even and $h=k-1$ (in which case $\mu_k$ and $\mu_h$
might even coincide).

\sss
Notice that, in particular, having $d$ eigenvalues as {\it independent}
parameters excludes the constant potential case $V\equiv$ constant (where, of course,
all eigenvalues are double: $\mu_{2j-1}=\mu_{2j}=j^2+V$). 
In fact, this case seems difficult to be handled by KAM 
approach even in the finite dimensional case. Such difficulty does not arise, 
instead, in the remarkable alternative approach developed by Craig, Wayne 
\cite{CW} and Bourgain \cite{B1}, \cite{B2}.

\sss
Equation (\ref{nonlinearwave}) may be rewritten as 
\begin{equation}\label{HH}
\dot u=v, \qquad \dot v +Au= f(u),
\end{equation}
which, as well known, may  be viewed as the (infinite dimensional) Hamiltonian
equations $\dot u = H_v$, $\dot v=- H_u$ associated to the Hamiltonian 
\begin{equation}
H=\frac 12( v,v)+\frac 12( Au,u)+\int_\T g(u)\ dx,
\end{equation}
where   $g$ is a primitive of $(-f)$
(with respect to the  $u$ variable) and  $( \cdot,\cdot)$ denotes the scalar product
in $L^2$.

\sss
As in \cite{P1}, we introduce  coordinates $q=(q_0, q_1, \cdots),
p=(p_0, p_1, \cdots)$  through the relations
\[u(x)=\sum_{n\in \N}\frac{q_n}{\sqrt{\lambda_n}}\phi_n(x),
\quad v=\sum_{n\in\N} \sqrt{\lambda_n}p_n\phi_n(x),
\]
where\footnote{Recall that, for simplicity, we assume that all eigenvalues $\mu_n$
are positive.}
$\lambda_n\equiv \sqrt{\mu_n}$. System (\ref{HH}) is then formally equivalent to
the lattice Hamiltonian equations
\begin{equation}\label{hameq} \dot q_n=\lambda_n p_n, 
\quad \dot p_n=-\lambda_nq_n- \frac{\partial G}{\partial q_n},
\quad G\equiv \int_\T
g(\sum_{n\in \N}\frac{q_n}{\sqrt{\lambda_n}}\phi_n)dx\ ,\end{equation}
corresponding to the Hamiltonian function  
$H= \sum_{n\in \N}\lambda_n(q_n^2+p_n^2)+ G(q)$.
Rather than discussing the above formal equivalence, we shall, following 
\cite{P1}, use the following elementary observation (proved in the Appendix):

\begin{Proposition}\label{prop}
Let $V$ be analytic (respectively, smooth), let $I$ be an interval and let
$$
t\in I\to (q(t), p(t))\equiv \Big(\{q_n(t)\}_{n\ge 0},\{p_n(t)\}_{n\ge 0}\Big)
$$
be an analytic (respectively, smooth\footnote{Regularity refers to the components
$q_n$ and $p_n$.}) solution of (\ref{hameq}) such
that
\begin{equation}\label{reg}
\sup_{t\in I} \sum_{n\in \Bbb N} \Big( |q_n(t)|+|p_n(t)|\Big) \ n^a\ e^{n
\rho}<\infty
\end{equation}
for some $\rho>0$ and $a=0$ (respectively, for $\rho=0$ and $a$ big enough).
Then
\[u(t, x)\equiv\sum_{n\in \N}\frac{q_n(t)}{\sqrt{\lambda_n}}\phi_n(x),\]
is an analytic (respectively, smooth) solution of (\ref{nonlinearwave}).
\end{Proposition}

\sss
Before invoking Theorem 1 we still need some manipulations.
We first switch to complex variables: 
$w_n=\frac{1}{\sqrt{2}}(q_n+{\rm i}p_n), \bar
w_n=\frac{1}{\sqrt{2}}(q_n-{\rm i}p_n)$.
Equations  (\ref{hameq}) read then
\begin{equation}\label{hamw} \dot w_n=-{\rm i}\lambda_n w_n- {\rm i}
\frac{\partial \tilde G}{\partial \bar w_n}, \quad
\dot{\bar w}_n={\rm i}\lambda_n\bar w_n+ {\rm i}
\frac{\partial \tilde G}{\partial w_n}, 
\end{equation}
where the perturbation ${\tilde G}$ is given by
\beq\label{tildeG}\tilde G(w)=\int_\T
g(\sum_{n\in \N}\frac{w_n+\bar w_n}{\sqrt{2 \lambda_n}}\phi_n)dx.
\eeq

\sss
Next we  
introduce standard  action-angle variables $(\theta, I)=((\theta_1,\cdots,
\theta_d),( I_1, \cdots, I_d))$ in the 
$(w_{n_1}, \cdots, w_{n_d},\bar w_{n_1}, \cdots, \bar w_{n_d} )$-space by
letting,
\[I_i= w_{n_i}\bar w_{n_i}, \quad i=1, \cdots, d,\]
so that the system (\ref{hamw}) becomes
\begin{eqnarray}\label{sysP}
\frac{d\theta_j}{dt}&=&\omega_j +  P_{I_j},\quad 
\frac {dI_j}{dt}=- P_{\theta_j},\quad j=1, \cdots, d,
\nonumber\\
\frac {dw_n}{dt}&=&-{\rm i}\lambda_nw_n-{\rm i}  P_{\bar w_n},\ \ \ 
\frac {d\bar w_n}{dt}={\rm i}\lambda_n\bar w_n +{\rm i} P_{w_n}, 
\ n\ne n_1, n_2,\cdots, n_d,
\end{eqnarray}
where $P$ is just $\tilde G$ with the
$(w_{n_1}, \cdots, w_{n_d},\bar w_{n_1}, \cdots, \bar w_{n_d} )$-variables 
expressed in terms of the $(\theta,I)$ variables and the {\sl frequencies}
$\omega=(\omega_1,...,\omega_d)$ coincide with the parameter $\xi$ introduced in
(\ref{xml}):
\begin{equation}\label{oml}
\omega_i\equiv \xi_i=\lambda_{n_i}\ .
\end{equation}
The Hamiltonian associated to (\ref{sysP}) (with respect to the
symplectic form
$dI\wedge d\theta+{\rm i}\sum_n dw_n \wedge d\bar w_n$) is given by
\begin{equation}\label{hamI}
H=\la\omega, I\ra+\sum_{n\ne n_1, \cdots, n_d}\lambda_nw_n\bar w_n+
P(\theta, I, w, \bar w, \xi),
\end{equation}

\ms\noindent
{\bf Remark} Actually, in place of $H$ in (\ref{hamI}) one should consider the
{\sl linearization} of $H$ around a given point $I_0$ and let $I$ vary in a 
small ball $B$ (of radius $0<s\ll |I_0|$) in the ``positive'' quadrant
$\{I_j>0\}$. In such a way the dependence of $H$ upon $I$ is obviously 
analytic. For notational convenience we shall however do not report explicitly
the dependence of $H$ on $I_0$.

\ms\noindent
Finally, to put the Hamiltonian in the form (\ref{hamH})
we couple the variables $(w_n, \bar w_n)$ corresponding
to ``closer" eigenvalues. More precisely, we let
$z_n=(w_{2n-1}, w_{2n}, \bar w_{2n-1}, 
\bar w_{2n})$ for large\footnote{Compare (A1).} $n$, say $n> \bar n> n_d$ and
denote by
$\displaystyle
z_0=(\{w_n\}_{ {0\le n\le \bar n}\atop{n\neq n_1,...,n_d}}, 
\{\bar w_n\}_{ {0\le n\le \bar n}\atop{n\neq n_1,...,n_d}})$
the remaining conjugated variables. The Hamiltonian  (\ref{hamI}) takes the form
\begin{equation}
\label{hamiltonian z}H
=\la\omega, I\ra+ \frac 12\sum_{n\in \N} \la A_n z_n, z_n\ra +
P(\theta, I, z, \xi),
\end{equation} 
where 
\begin{eqnarray}
A_n&=&{\rm Diag}(\lambda_{2n-1},\lambda_{2n},\lambda_{2n-1},\lambda_{2n}) \pmatrix{ 0
& I_2 \cr I_2 & 0 \cr}\nonumber\\
&=& \lambda_{2n} \pmatrix{ 0 & I_2 \cr I_2 & 0 \cr} +
\pmatrix{0&0&\lambda_{2n-1}-\lambda_{2n}&0\cr
0&0&0&0\cr
\lambda_{2n-1}-\lambda_{2n}&0&0&0\cr0&0&0&0\cr}
\ ,\nonumber
\end{eqnarray}
for $n>n_d$, while $A_0={\rm Diag}(\{\lambda_n\},$
$  \{\lambda_n\}; 1\le n\le n_d, n\ne n_1, \cdots, 
n_d) \pmatrix{ 0 & I_{d_0} \cr  I_{d_0} & 0 \cr}$ with $d_0=\bar n +1 -d$.

\sss
The perturbation $P$ in (\ref{hamiltonian z}) has the following (nice) regularity
property.
\begin{Lemma}\label{regularity}
Suppose that  $V$ is real analytic in $x$ (respectively, belongs to the 
Sobolev space $H^k(\T)$ for some $k\in \N$). 
Then for small enough $\rho>0$ (respectively, $a>0$), $r>0$ and $s>0$
one has
\beq\|X_P\|_{\!{}_{D_{a, \rho}(r,s), \cal O}}^{a+1/2,\rho}= 
O(|z|_{a,\rho}^2) \ ;
\label{Pz}
\eeq
here the parameter $a$ is taken to be 0 (respectively, the parameter $\rho$ is taken
to be 0).
\end{Lemma}
A proof of this lemma is given in the Appendix. In fact, $X_P$ 
is even more ``regular" (a fact, however, not needed in what follows):
(\ref{Pz}) holds with 1 in place of 1/2. 

\ms
The Hamiltonian (\ref{hamiltonian z}) is seen to satisfy all the assumptions of
Theorem \ref{KAM} with: 
$d_n=4,n\ge 1;$  $d_0=\bar n +1 -d$; $\bar d=\max\{d_0,
4\}$; $b=1$; $\delta=2$; 
$\delta_1$ choosen as in (\ref{del1});
$\bar a-a=\frac 12$.
Thus Theorem \ref{KAM} yields the following

\begin{Theorem}
Consider a family of 1D nonlinear wave equation (\ref{nonlinearwave})
parameterized by $\xi\equiv \omega\in \Cal O$ as above with  $V(\cdot, \xi)$ 
real--analytic (respectively, smooth).
Then for any $0<\gamma\ll 1$, there is a subset $\Cal O_\gamma$ of $\Cal O$ with 
${\rm meas}(\Cal O\backslash \Cal O_\gamma)\to 0$ as $\gamma\to 0$,
such that 
${\rm (\ref{nonlinearwave})}_{\xi\in \Cal
O_\gamma}$ has a family of small-amplitude (proportional to 
some power of $\gamma$), analytic (respectively, smooth)
quasi-periodic solutions of the form
\[ u(t, x)=\sum_n u_n(\omega_{1}'t, \cdots, \omega_{d}' t)
\phi_n(x)\]
where $u_n: \T^d\to \R $ and $\omega_{1}',\cdots, \omega_{d}'$
are close to $\omega_{1},\cdots, \omega_{d}$.
\end{Theorem} 

\noindent
{\bf Remark}
As mentioned above, our KAM theorem (which applies only
to the case that not all the eigenvalues are multiple\footnote{
Recall that we require that the torus frequencies are independent parameters.}
and under the hypothesis that all $\mu_n$'s are
positive) implies that the quasi-periodic solutions obtained  are {\it linearly
stable}. In the case that all the eigenvalues are double (as in the constant
potential case), one should not expect linear stability (see the example given by
Craig, Kuksin and  Wayne \cite{CKW}).
We also notice that, essentially with only notational changes, the proof of the above
theorem goes through in the case that some of the eigenvalues are negative.

\section{ KAM step}

Theorem 1 will be proved by a KAM iteration which involves an infinite
sequence of change of variables.

\sss
At each step of the KAM scheme,  we consider a Hamiltonian vector
field with  

$$ H_\nu=N_\nu+P_\nu,$$ 
where $N_\nu$ is an ``integrable normal form" and
$P_{\nu}$ is
defined in some set of the form\footnote{Recall the notations from Section 2.} 
$D(s_\nu, r_\nu)\times \Cal O_{\nu} $.

\sss
We then construct a map\footnote{Recall that the parameters $a$, $\rho$ and $\bar a$
are fixed throughout the proof and are therefore omitted in the notations.}
$$\Phi_\nu: D (s_{\nu+1}, r_{\nu+1})\times\Cal O_{\nu+1}
\subset  D(r_\nu, s_\nu)\times\Cal O_\nu \to D(r_\nu, s_\nu)
\times\Cal O_{\nu}
$$
so that the vector field $X_{H_\nu
\circ\Phi_\nu}$ defined on  $D(r_{\nu+1}, s_{\nu+1})$
satisfies
\[\|X_{H_\nu\circ\Phi_\nu}-X_{N_{\nu+1}}\|_{r_{\nu+1}, s_{\nu+1},
\Cal O_{\nu+1}}\le \epsilon_{\nu}^\kappa\]
with some new normal form $N_{\nu +1}$ and for some fixed $\nu$-independent
constant $\kappa>1$.

\sss
To simplify notations, in what follows, the quantities without subscripts refer to
quantities at the  $\nu^{\rm th}$ step, while the quantities with subscripts $+$
denotes the corresponding quantities at the  $(\nu+1)^{\rm th}$ step. 
Let us then consider the Hamiltonian
\begin{equation} H=N+P\equiv e+ \la\omega,I\ra
+\frac 12\sum_{n\in \N} \la A_nz_n, z_n\ra
+P,
\label{3.1}
\end{equation}
defined in $D(r, s)\times\Cal O$; teh $A_n$'s are symmetric matrices. We
assume that  
$\xi\in \Cal O$ satisfies\footnote
{
The tensor product (or direct product) of two $m\times n$, $k\times l$
matrices $A=(a_{ij}), B$ 
is a $(mk)\times (nl)$ matrix 
defined by
\[ A\otimes B=(a_{ij}B)= \left(\begin{array}{ccc}
a_{11}B & \cdots & a_{1n}B\\
     \cdots & \cdots & \cdots \\
a_{n1}B & \cdots & a_{mn}B
\end{array}\right)
.\]
$\|\cdot \|$ for matrix denotes the operator norm, i.e.,
 $\|M\|=\sup_{|y|=1}|My|.$
Recall that $\omega$ and the $A_i$'s depend on $\xi$.
}
(for a suitable $\tau>0$ to be specified later)  

\begin{eqnarray}\label{ODC}
& &|\la k,\omega\ra^{-1} |< \frac{|k|^\tau}{\gamma} ,\;\;\;\;
\|(\la k,\omega\ra I_{d_i}+ 
{ A_iJ_{d_i}})^{-1}\|<(\frac{|k|^\tau}{\gamma})^{\bar d}
\nonumber\\
& &\|(\la k, \omega\ra  I_{d_id_j}+(A_iJ_{d_i})
\otimes I_{d_j}-I_{d_i}\otimes (J_{d_j}A_j))^{-1}\|
<(\frac{|k|^\tau}{\gamma})^{\bar d^2},
\end{eqnarray}

\sss
We also assume that 
\begin{equation}\label{aijM}
\max_{|p|\le \bar d^2}\|\frac{\partial^p A_n}{\partial\xi^p}\|\le 
L,
\end{equation}
on $\Cal O$,  and 
\begin{equation}\|X_P\|_{ r,s,\Cal O}
\le \epsilon.
\label{3.3}
\end{equation}
We now let $0<r_+<r$, and define
 \begin{equation}\label{+} 
 s_+=\frac 12s\epsilon^{\frac 13}, \quad  
\epsilon_+=\gammaaa\Gamma(r-r_+)
\epsilon^{\frac 43}, 
\label{3.4-}
\end{equation}
where 
$$\Gamma(t)\equiv \sup_{u\ge 1} u^{\rm c}  e^{-\frac 14ut}
\sim t^{-{\rm c}}.$$
for $t>0$. Here and later, the letter $c$ denotes suitable (possibly different)
constants that do not depend on the iteration step\footnote{Actually, here 
$c=\bar d^4\tau+\bar d^2\tau+\bar d^2+1$.}.  

\sss
We now describe how  to construct a set $\Cal O_+\subset \Cal O$ and
a change of variables  
 $\Phi: D_+\times\Cal O_+=D(r_+, s_+)\times \Cal O_+\to D(r,s)\times \Cal O$, 
such that the transformed Hamiltonian 
$H_+=N_++P_+\equiv H\circ \Phi$   satisfies 
all the above iterative assumptions with new parameters
$s_+, \epsilon_+, r_+,  \gamma_+, L_+$ and with  $\xi\in \Cal O_+$.

\ms
\subsection{Solving the linearized equation}

\sss 
Expand $P$ into the  Fourier-Taylor series
$$P=\sum_{k,l,\alpha} P_{kl\alpha}\kth I^lz^\alpha$$
where $k\in \Z^d, l\in \N^d$ and  $\alpha\in \otimes_{n\in {\Bbb N}}\Bbb N^{d_n}$ 
with finite many non-vanishing  components.

\sss
Let  $R$ be the truncation of $P$ given by
\begin{eqnarray}
R(\theta,I,z)&\equiv &P_0+P_1+P_2
\equiv \sum_{k,|l|\le 1} P_{kl0}\kth I^l\nonumber\\
& &+\sum_{k,|\alpha|=1} P_{k0\alpha}\kth z^\alpha+\sum_{
k,|\alpha|=2}
 P_{k0\alpha} \kth z^\alpha,\label{Pi}\end{eqnarray}
with 
\[2|l|+|\alpha|=2\sum_{j=1, \cdots, d}
l_j+
\sum_{j\in \N} |\alpha_j| \le 2.\] 

\sss
It is convenient to rewrite $R$ as follows
\begin{eqnarray}\label{R}
R(\theta, I, z)&=& \sum_{k,|l|\le 1} P_{kl0}\kth I^l\nonumber\\
&+&\sum_{k,i} \la R^{k}_i,z_i\ra \kth +\sum_{k,i,j}
 \la{ R}^k_{ji}z_i, z_j\ra\kth,
\end{eqnarray}
where $R^k_i,R^k_{ji}$ are respectively the $d_i$ vector and $(d_j\times d_i)$
matrix defined by
\begin{equation}\label{R_k}
R^{k}_i=\int\frac{\partial P}{\partial z_i}e^{-{\rm i}\langle k,\theta
\rangle}d\theta |_{z=0,I=0}, \ \ \
R^k_{ji}=
\frac {1+\delta_i^j}2\int\frac{\partial^2 P}{\partial z_j\partial z_i}
e^{-{\rm i}\langle k,\theta
\rangle}d\theta|_{z=0, I=0}.
\end{equation}
 Note that  $R^k_{ij}=(R^k_{ji})^T$.

\sss
Rewrite $H$ as 
$ H=N+R+(P-R)$. By the choice of $s_+$ in (\ref{+}) and by the definition of 
the norms, it follows immediately that

\begin{equation}
\label{P-R}
\|X_R\|_{r,s,\Cal O}\le
\|X_P\|_{r,s,\Cal O}\le\epsilon.
\end{equation} 
Moreover $s_+, \epsilon_+$ are such that, in a smaller domain $D(r, s_+)$, we have
\begin{equation}
\label{P-R+}
\|X_{P-R}\|_ {r, s_+}  \lep \epsilon_+. 
\end{equation}

\sss
Then we  look for  a special $F$, 
defined in  domain $D_+=D(r_+, s_+),$
such that the time one  map $\phi^1_F$ of the Hamiltonian vector field  $X_F$ 
defines a map from $D_+\to D$ and transforms
$H$ into $H_+$.

\sss
More precisely, by second order Taylor formula, we have
\begin{eqnarray}
H\circ \phi^1_F &=&(N+R)\circ \phi_F^1+(P-R)\circ \phi^1_F\nonumber\\
&=& N+ \{N,F\}+R\nonumber\\
& &\ \ \ \ +\frac 12\int_0^1ds\int_0^s\{\{N+R,F\},F\}\circ \phi_F^{t}dt+\{R,F\}
+(P-R)\circ \phi^1_F.
\nonumber\\
&=& N_++P_+\nonumber\\
& & + \{N,F\}+R-P_{000}-\la\omega', I\ra-\sum_{n\in \N}
\la R^0_{nn}z_n, z_n\ra,
\label{2.3}
\end{eqnarray}
where 
\[ 
\omega'= \int\frac{\partial P}{\partial I}d\theta|_{I=0, z=0},\quad
R^0_{nn}=\int \frac{\partial^2 P}{\partial z_n^2}d\theta|_{I=0, z=0},
\]
$$N_+= N+P_{000}+\la\omega', I\ra+
\sum_{n\in \N}\la R^0_{nn}z_n, z_n\ra
 $$
$$P_+=\frac 12\int_0^1ds\int_0^s\{\{N+R,F\},F\}\circ X_F^{t}dt+\{R,F\}
+(P-R)\circ \phi^1_F.$$

\ms
We shall find  a function $F$ of the form
\begin{eqnarray}\label{F}
F(\theta, I,z)&=& F_0+ F_1+F_2=\sum_{|l|\le 1,|k|\ne 0} 
F_{kl0}\kth
I^l+
\sum_{i\in \N} \la F^k_i,z_i\ra\kth \nonumber\\
& &+\sum_{|k|+|i-j|\ne 0} \la { F}^k_{ji}z_i, z_j\ra\kth,
\label{3.14}
\end{eqnarray}
satisfying the equation 
\begin{equation}\label{hom} 
\{N,F\}+R-
P_{000}-\la\omega', I\ra -\sum_{n\in \N}\la R^0_{nn}z_n, z_n\ra =0.
\end{equation}

\begin{Lemma}\label{Fdef} Equation (\ref{hom}) is equivalent to
\begin{eqnarray}\label{Feq}
F_{kl0}&=& ({\rm i}\la k,\omega\ra )^{-1} P_{kl0}, \quad \ k\ne 0, |l|\le 1,
\nonumber\\
(\la k,\omega\ra I_{d_i} &+& { A_{d_i}J_{d_i}})F^k_i={\rm i} R^k_i,\nonumber\\ 
(\la k,\omega\ra I_{d_i}&+& A_{d_i}J_{d_i})F^k_{ij}- F^k_{ij}
( J_{d_j}A_j)={\rm i} R^k_{ij},\ \, |k|+|i-j|
\ne 0.
\end{eqnarray} 
\end{Lemma}

\Proof Inserting $F$, defined in (\ref{F}), into (\ref{hom})  one sees that 
(\ref{hom}) is equivalent to 
 the following equations\footnote{Recall the definition of $P_i$ in
(\ref{Pi}).}
\begin{eqnarray}
\label{Fcoefficents}
& &\{N, F_0\}+ P_0-\la\omega', I\ra=0, \nonumber\\
& & \{N, F_{1}\}+ P_{1}=0,\nonumber\\ 
& &\{N, F_{2}\}+ { P}_{2}-\sum_{n\in Z}\la R^0_{nn}z_n, z_n\ra=0.
\end{eqnarray}
The first equation in (\ref{Fcoefficents}) is obviously equivalent, by
comparing the coefficients, to the first equation in (\ref{Feq}).
To solve $\{N, F_1\}+  P_1=0$, we note that\footnote{Recall the definition
of $N$ in (\ref{3.1}).}
\begin{eqnarray}
\{N, F_1\}&=& \la \partial_I N,\partial_\theta F_1\ra
+ \la \nabla_zN, { J}\nabla_zF_1\ra\nonumber\\
&=& \la \partial_I N,\partial_\theta F_1\ra
+\sum_{i} \la\nabla_{z_i}N, {{\rm i} J_{d_i}}\nabla_{z_i}F_1\ra
\nonumber\\
&=&{\rm i}\sum_{k,i} 
( \la \;\ko F^k_i,z_i\ra+\la A_iz_i, J_{d_i}F^k_i\ra) \kth
\nonumber\\
&=&{\rm i}\sum_{k,i}  \la (\ko I_{d_i}+A_iJ_{d_i})F^k_i, z_i\ra   
\kth.
\end{eqnarray}
It follows that
$F^k_i$ are determined by the  
linear algebraic system
$${\rm i}(\la k,\omega\ra { I_{d_i}}+ { A_iJ_{d_i} })F^k_i+ R^k_i=0,
\quad i\in \N, k\in \Z^d$$

\sss
Similarly, from
\begin{eqnarray}
\{N, F_2\}
&=& \la\partial_IN, \partial_\theta F_2\ra 
+\sum_{i} \la\nabla_{z_i}N, { {\rm i}J_{d_i}}\nabla_{z_i}F_2\ra\nonumber\\
&=&{\rm i}\sum_{ |k|+|i-j|\ne 0} (\la\;\la k,\omega\ra F^k_{ji}z_i,z_j\ra
+\la A_iz_i, J_{d_i}(F^k_{ji})^Tz_j\ra +\la A_jz_j, J_{d_j}F^k_{ji}z_i\ra)
\kth\nonumber\\
&=&{\rm i}\sum_{ |k|+|i-j|\ne 0} (\la\;\ko { F}^k_{ji}z_i,z_j\ra
+\la (A_jJ_{d_j}F^k_{ji}-F^k_{ji}J_{d_i}A_i)z_i,z_j\ra)\kth\nonumber\\
&=&{\rm i}\sum_{ |k|+|i-j|\ne 0}
 \la (\la k,\omega\ra F^k_{ji}+A_{j}J_{d_j}F^k_{ji}-
F^k_{ji}J_{d_i}A_i)z_i,z_j\ra
\kth
\end{eqnarray}  
it follows that, $F^k_{ji}$ is determined by the following
 matrix equation
\begin{equation}\label{matrixeq}
(\la k,\omega\ra I_{d_j}+ A_jJ_{d_j})F^k_{ji}- F^k_{ji}(J_{d_i}A_i) 
={\rm i}R^k_{ji}, \quad |k|+
|i-j|\ne 0
\end{equation}
where $ F^k_{ji},  R^k_{ji}$ are  $d_j\times d_i$ matrices defined in 
(\ref{F}) and (\ref{R}). Exchanging $i, j$ we get the third equation in   
(\ref{Feq}).
\qed

\ms
The first two equations in (\ref{Feq}) are immediately solved in view of
(\ref{ODC}).
In order to solve the third equation in (\ref{Feq}), we need the following 
elementary algebraic result from  matrix theory.
\begin{Lemma}\label{matrixsolution}
Let $A, B, C$ be respectively $n\times n, m\times m, n\times m$ matrices,
and let $X$ be an $n\times m$ unknown matrix.
The matrix equation 
\beq\label{meq} AX-XB=C,\eeq
is solvable if and only if  $I_m\otimes A-B\otimes I_n$ is nonsingular. Moreover,
\[ \|X\|\le  \|( I_m\otimes A-B\otimes I_n )^{-1}\|\cdot \|C\|.\]
\end{Lemma}

\sss
In fact, the matrix equation (\ref{meq}) is equivalent to the
 (bigger) vector equation given by 
$(I\otimes A-B\otimes I)X'=C'$ where $X', C'$ are vectors whose elements
are just the list (row by row) of the entries of $X$ and $C$. 
For a detailed proof we refer the reader to the Appendix in \cite{Y} or
\cite{Lan}, p. 256. 

\sss
\noindent
{\bf Remark}. Taking the transpose of the third equation in (\ref{Feq}),
one sees  that $(F^k_{ij})^T$ satisfies the same equation of $F^k_{ji}$. Then
(by the uniqueness of the solution) it follows that
$F^k_{ji}=(F^k_{ij})^T$.

\ms
\subsection{Estimates on the coordinate transformation}

\sss
We proceed to estimate $X_F$ and $\Phi_F^1$. We start with the following

\begin{Lemma}\label{Festimate} 
Let $D_i=D(\frac i4 s, r_++\frac{i}4 (r-r_+))$,
$0 <i \le 4$. Then
\begin{equation}
\|X_F\|_{D_3, \Cal O}\lep \gammaa\Gamma(r-r_+)  \epsilon.
\end{equation}
\end{Lemma}
\Proof
By  (\ref{ODC}), Lemma \ref{Fdef} and Lemmata \ref{extension1}, 
{\ref{extension2} in the Appendix,   we have
\begin{eqnarray}
|F_{kl0}|_{\Cal O}&\le & |\ko|^{-1} |P_{kl}|
\lep \gammaa\kk
e^{-|k|(r-r_+)}\epsilon
s^{2-2|l|}, \quad  k\ne 0,\nonumber\\
\|F^k_i\|_{\Cal O}&=& \|(\ko {I_{d_i}} + {A_iJ_{d_i}})^{-1}R^k_i\|\le
\|(\ko I_{d_i}+ A_iJ_{d_i})^{-1}\|\cdot \|R^k_i\|\nonumber\\
&<&  {\rm c}\ \gammaa\kk |R^k_i| ,\nonumber\\
\|F^k_{ij}\|_{\Cal O}&\le & 
\|(\ko {I_{d_id_j}} +(A_iJ_{d_i})\otimes I_{d_j}-  
I_{d_i}\otimes (J_{d_j}A_j))^{-1}\|
\cdot \|R^k_{ij}\|\nonumber\\
&<& {\rm c}\  \gammaa\kk\|R_{ij}^k\|, \, \,|k|+|i-j|\ne 0.
\end{eqnarray}
Where $\|\cdot\|_{\Cal O}$ for matrix is similar to (\ref{Falpha}).

\sss
It follows that
\begin{eqnarray}
\frac 1{s^2}\|F_\theta\|_{D_2,\Cal O}
&\le & \frac 1{s^2}(\sum |f_{kl0}|\cdot|I^l|\cdot |k|\cdot |\kth|
+\sum |F^k_i| \cdot|z_i|\cdot |k|\cdot |\kth|\nonumber\\
&+&\sum \|F^k_{ij}\|\cdot |z_i|\cdot|z_j|\cdot |k|\cdot |\kth|)
\nonumber\\
& \lep &
 \gammaa
\Gamma(r-r_+)\|X_R\|\nonumber\\
&\lep &\gammaa \Gamma(r-r_+)\epsilon,
\end{eqnarray}
where $\Gamma(r-r_+)=\sup_{k} \kk e^{-|k|\frac 14(r-r_+)}.  $
\sss
Similarly,
\[ \|F_I\|_{D_2,\Cal O}=\sum_{|l|\le 1}|F_{kl0}|\cdot |\kth|
\lep  \gammaa \Gamma(r-r_+)\epsilon.
\]

\sss
Now we estimate $\|X_{F^1}\|_{D_{2}, \Cal O}$.
Note that
\begin{eqnarray} \|F^1_{z_i}\|_{D_{2}, \Cal O}
&=&\|\sum_kF^k_i e^{-i<k,\theta>}\|_{D_{2}, \Cal O}
\nonumber\\
&\lep& \gammaa\Gamma \sum_{k,i}|R^k_i|e^{|k|r}
\lep \gammaa\Gamma \|\frac{\partial P_1}{\partial z_i}\|.
\end{eqnarray}
It follows that 
\begin{eqnarray}
\|X_{F^1}\|_{D_{2}, \Cal O}&\lep&\sum_{i\in \N}
\|F^1_{z_i}\|_{D_{2}, \Cal O}i^ae^{i\rho}\nonumber\\
&\lep &\gammaa\Gamma \sum_{i\in \N}
\|\frac{\partial P_1}{\partial z_i}\|i^ae^{i\rho}
\lep \gammaa
\Gamma \epsilon,\nonumber
\end{eqnarray}
 by the definition of the weighted norm.

\sss
Note that
\begin{eqnarray} \|F^2_{z_i}\|_{D_{2}, \Cal O}
&=&\|\sum_{k,j} (F^k_{ij}+(F^k_{ij})^T)z_j \kth\|_{D_{2}, \Cal O}
\nonumber\\
&\lep&  \gammaa\Gamma \|\frac{\partial P_2}{\partial z_i}\|.
\end{eqnarray}

\sss
Similarly, we have
\begin{equation}\label{3.16}
 \|X_{F^2}\|_{D_{2}, \Cal O}\lep 
\gammaa \Gamma \epsilon.
\end{equation}
The conclusion of the lemma follows from the above estimates.
\qed

\sss
In the next lemma, we give some estimates for $\phi_F^t$.
The following formula (\ref{3.32}) will be used to prove that our coordinate
transformations is well defined.
 (\ref{3.33}) is for proving the convergence of the iteration.

\sss
\begin{Lemma}\label{(3.5)}
Let  $\eta=\epsilon^{\frac 13}, D_{\frac i2\eta}=
D(r_++\frac {i-1}2(r-r_+),\frac{i}2 \eta s), i=1,2$. 
We then have
\begin{equation} 
\phi_F^t:  D_{\frac 12\eta}
\to  D_{\eta} ,\ \ \ 0\le t\le 1,
\label{3.32}
\end{equation}
if $\epsilon\ll (\frac 12\gammaa \Gamma^{-1})^
{\frac 32}$.  Moreover,
\begin{equation}
\|D\phi_F^1-Id\|_{D_{\frac 12\eta}}\lep \gammaa\Gamma\epsilon.
\label{3.33}
\end{equation} 
\end{Lemma}
\Proof
Let 
$$\|D^mF\|_{D,\Cal O}
=\max \{ |\frac{\partial^{|i|+|l|+p}}{\partial  \theta^{i}\partial I^{l}
\partial  z^\alpha} F|_{D, \Cal O}, |i|+|l|+|\alpha|=m\ge 2\}.$$

\sss
Note that  $F$ is polynomial in $I$ of order 1, in $z$ of order 2.
 From\footnote{Recall the definition of the weighted norm in (\ref{weightednorm}).}
 (\ref{3.16}) and the Cauchy inequality,
 it  follows 
that
 \begin{equation}\|D^mF\|_{D_1, \Cal O }\lep
 \gammaa \Gamma\epsilon,\label{3.17}
\end{equation} 
for any $m\ge 2$.

\sss
To get the estimates for $\phi_F^t$, we start from the integral equation, 
$$\phi_F^t=id+\int_0^tX_F\circ \phi_F^s\,ds$$

$\phi_F^t: D_{\frac 12\eta}
\to  D_{\eta} ,\ \ \ 0\le t\le 1$, follows directly from (\ref{3.17}).
 Since 
$$D\phi_F^1=Id+\int_0^1(DX_F) D\phi_F^s\,ds=
Id+\int_0^1J(D^2F) D\phi_F^s\,ds,$$
it follows that 
\begin{equation}\|D\phi_F^1-Id\|\le 2\|D^2F\|\lep 
\gammaa\Gamma \epsilon.
\label{3.36}
\end{equation}
The estimates of second order derivative $D^2\phi_F^1$ follows from
(\ref{3.17}). 
\qed

\ms
\subsection{Estimates for the new normal form}

\sss
The map $\phi_F^1$ defined above transforms $H$ into
$H_+=N_+ +  P_+$(see (\ref{2.3}) and (\ref{hom})) 
with 
\begin{equation}  
N_+= e_++\la \omega_+, y\ra+\frac 12\sum_{i\in {\Bbb Z}_+}\la A^+_iz_i,z_i\ra
\end{equation}
where
\begin{equation} e_+=e+P_{000},\ 
\omega_+=\omega+P_{0l0} (|l|=1),\  
 A_i^+=A_i+2 R^0_{ii}.
\end{equation}

\ms
Now we prove that $N_+$ shares the same properties with $N$.
By the regularity of $X_P$ and by Cauchy estimates, we have 
\begin{equation}\label{Rest}
|\omega_{+}-\omega|<\epsilon, \quad
\|R^{0}_{ii}\|<\epsilon i^{-\delta}
\end{equation} 
with $\delta=\bar a -a>0 $.
It follows that
\begin{eqnarray} & & \|(A_i^+)^{-1}\|\le  \frac
{\|A_i^{-1}\|}{1-2\|A_i^{-1}R^0_{ii}\|} \le 2\|A_i^{-1}\|, 
\nonumber\\
& & \|(\la k,\omega+P_{0l00}\ra I_{d_i}-{J_{d_i}A_i^+})^{-1}\|
\le 
\frac{\|(\ko I_{d_i}+{A_iJ_{d_i}})^{-1}\|}
{1- \|(\ko I_{d_i}+{A_iJ_{d_i}})^{-1}\|\epsilon}
\le (\frac{ |k|^\tau}{ \gamma_+})^{\bar d},
\label{3.2+}
\end{eqnarray}
provided $|k|^{\bar d\tau}\epsilon\lep (\gamma^{\bar d}-\gamma_+^{\bar d}) $.

\sss
Similarly, we have
\begin{equation}\label{4.30}
\|(\la k,\omega+P_{0l00}\ra I_{d_id_j}+
(A_i^+ J_{d_i})\otimes I_{d_j}- I_{d_i}\otimes
(J_{d_j}A_j^+))^{-1}\|\le
 (\frac {|k|^\tau}{\gamma_+})^{\bar d^2},
\end{equation}
provided $|k|^{\bar d^2\tau}\epsilon\lep (\gamma^{\bar d^2}-\gamma_+^{\bar d^2}) $.
This means that in the next KAM step, small denominator conditions are 
automatically satisfied for 
$|k|<K$ where  $K^{\bar d^2\tau}\epsilon\lep (\gamma^{\bar d^2}-\gamma_+^{4\bar
d^2}) $.

\sss
The following bounds wil be used later for the measure estimates.

\begin{equation}\label{a+}
|\frac{\partial^l(\omega_+-\omega)}
{ \partial\xi^l}|_{\Cal O}\le \epsilon,\quad
  |\frac{\partial^l(A_{i}^+-A_{i})}
{ \partial\xi^l}|
_{\Cal O} \lep \epsilon i^{-\delta},\end{equation}
for $|l|\le \bar d^2$ by the definition of the norm.

\subsection{ Estimates for the new perturbation}

To complete the KAM step we have to estimate the new error term.

\sss
By the definition of $\phi^1_F$  and 
Lemma \ref{(3.5)},
$$ H\circ \phi^1_F
=N_++ P_+ 
$$
is well defined in  $D_{\frac 12\eta}.$
 Moreover, we have the following estimates
\begin{eqnarray}
\|X_{P_+}\|_{D_{\frac 12\eta}}
&=& \|X_{\int_0^1dt\int_0^s\{\{N+R, F\},F\}\circ  \phi^s_F 
+\{R,F\}+(P-R)\circ\phi^1_F}\|_{D_{\frac 12\eta}}
\nonumber \\
&\le&\|X_{(\int_0^1dt\int_0^t\{\{N+R, F\},F\}\circ  
\phi^s_F }\|_{D_{\frac 12\eta}}
+\|X_{(P-R)\circ\phi^1_F}\|_{D_{\frac 12\eta}}
\nonumber \\
&\le &\|X_{\{\{N+R, F\},F\}}\|_{D_{\eta}}
+\|X_{P-R}\|_{D_{\eta}}\nonumber\\
&\lep& \gammaaa \Gamma^2
\epsilon^{\frac 43}\lep \epsilon_+
\label{3.37}
\end{eqnarray}
by (\ref{P-R}) and Lemma \ref{FG}.

\sss
Thus, there exists a big constant $c$, 
independent of iteration steps, such that
\begin{equation}\label{P+}
\|X_{P_+}\|_{r_+, s_+}=\|X_{P_+}\|_{D_{\frac 12\eta}}^{\bar a, \rho}
\le c \gammaaa\Gamma^2\eta\epsilon=c\epsilon_+.
\end{equation}

\sss
The KAM step is now completed.

\section{Iteration Lemma and Convergence}

For any given $s, \epsilon, r, \gamma$, we define, for all $\nu\ge 1$, the
following sequences
\begin{eqnarray}\label{iterationconstants}
r_{\nu} &=&r(1- \sum_{i=2}^{\nu+1 }2^{-i}),\nonumber\\
\epsilon_{\nu} &=& c\gamma_{\nu}^{\gammab}\Gamma(r_{\nu-1}-r_{\nu})^2
\epsilon_{\nu-1}^{\frac 43}.\nonumber\\
 \gamma_{\nu} &=&\gamma(1-\sum_{i=2}^{\nu+1 }2^{-i})\nonumber\\ 
\eta_{\nu}& =& \frac 12\epsilon_{\nu}^{\frac 13},\quad
 L_{\nu} = L_{\nu-1}+\epsilon_{\nu-1}, \nonumber\\
s_{\nu} &=& \frac 12\eta_{\nu-1} s_{\nu-1}= 2^{-\nu}
(\prod_{i=0}^{\nu-1}\epsilon_i)^{\frac 13}s_0,\nonumber\\
 K_{\nu} &=& \frac c2\left(\epsilon_{\nu-1}^{-1}
(\gamma_{\nu-1}^{\bar d^2}
-\gamma_{\nu}^{\bar d^2})\right)^{\frac1{\bar d^2\tau}},\nonumber\\
D_{\nu} &=& D_{a,\rho}(r_{\nu}, s_{\nu}),
\end{eqnarray} 
where $c$ is the constant in (\ref{P+}). The parameters
$r_0, \epsilon_0, \gamma_0, L_0, s_0, K_0$ are defined respectively to be
$r, \epsilon, \gamma, L, s, 1$.

\ms
Note that 
$$
\Psi(r) =\prod_{i=1}^\infty [\Gamma(r_{i-1}-r_i)]^{2(\frac 34)^{i}},$$
is a well defined finite function of $r$.

\ms
\subsection{Iteration Lemma}

\sss
The analysis of the preceeeding section can be summarized as follows.

\begin
{Lemma}\label{iteration} Suppose that  $\epsilon_0=\epsilon(d, \bar
d, \delta, \delta_1,
\bar a-a, L, \tau, \gamma)$
 is small enough. 
Then the following holds for all $\nu\ge 0$.
Let 
$$N_\nu=e_\nu+\la\omega_\nu(\xi),I\ra
+\sum_{i\in \N}\la A^\nu_i(\xi) z_i, z_i\ra
,$$ 
 be a normal form with parameters $\xi$ satisfying
\begin{eqnarray}
& &|\la k,\omega_\nu\ra^{-1} |<\frac{|k|^{\tau}}{\gamma_\nu }, \;\;\;
\|({\rm i}\la k,\omega_\nu\ra { I_{d_i}}+{ A^\nu_iJ_{d_i} })^{-1}\|<
(\frac {|k|^{\tau}}{\gamma_\nu })^{\bar d}
\nonumber\\
& &\|({\rm i}\la k,\omega_\nu\ra I_{d_id_j}+(A_i^\nu J_{d_i} )\otimes I_{d_j}
-I_{d_i}\otimes (J_{d_j}A_j^\nu )
)^{-1}\|
<(\frac{|k|^{\tau}}{\gamma_\nu })^{\bar d^2}
\end{eqnarray}
on a closed set $\Cal O_\nu$ of $R^n$ for 
all $k\ne 0, i, j\in \Z$.
Moreover,
suppose that $\omega_\nu(\xi), A_{i}^\nu(\xi)$ are $C^{\bar d^2}$
smooth  and satisfy
$$|\frac{\partial^{\bar d^2}(\omega_\nu-\omega_{\nu-1})}
{\partial\xi^{\bar d^2}}|
\le \epsilon_{\nu-1},\
|\frac{\partial^{\bar d^2}(A_{i}^\nu-A_{i}^{\nu-1})}
{\partial\xi^{\bar d^2}}
|\le \epsilon_{\nu-1} i^{-\delta},$$  
on  $\Cal O_\nu$ in Whitney's sense.

\sss
Finally, assume that   $$\|X_{P_\nu}\|_{D_\nu,\Cal O_{\nu}}^{\bar a,\rho}
\le \epsilon_{\nu}.$$  
 Then,  there 
is a subset $\Cal O_{\nu+1}\subset \Cal O_{\nu}$,
$$\Cal O_{\nu+1} =\Cal O_{\nu}- 
 \cup_{ |k|\ge K_{\nu+1}} {\Cal R}^{\nu+1} _{kij}(\gamma_{\nu})$$
where
\begin{displaymath}
\Cal R_{kij}^{\nu+1}(\gamma_{\nu+1})=  
\left\{\xi\in \Cal O_\nu: 
|\; \stackrel{\scriptstyle |\la k,\omega_{\nu+1}\ra^{-1}|>\frac 
{|k|^{\tau}}{\gamma_{\nu}},\quad
 \|(\la k,\omega_{\nu}\ra I_{2m}+
(A_i^{\nu+1}J_{d_i})^{-1} \|\ge (\frac 
{|k|^{\tau}}{\gamma_{\nu}})^{\bar d},\, \mbox{or}}
 {\scriptstyle \|(\la  k,\omega_{\nu+1}>I_{d_id_j}+(A_j^{\nu+1}J_{d_i})
\otimes I_{d_j}-I_{d_i}\otimes
(J_{d_j}A_j^{\nu+1} ))^{-1}\|
> (\frac
{|k|^{\tau}}{\gamma_{\nu}}
})^{\bar d^2}\;\right\}, 
\end{displaymath}
with $\omega_{\nu+1}= \omega_\nu+P_{0l0}^\nu$,
and a symplectic  change of variables 
\begin{equation}
\Phi_\nu: D_{\nu+1}\times \Cal O_{\nu+1} \to D_\nu,
\label{4.1}
\end{equation}
such that 
$H_{\nu+1}=H_\nu \circ\Phi_\nu$, defined on
 $D_{\nu+1}\times \Cal O_{\nu+1} $,
 has the form
\begin{equation}H_{\nu+1}=e_{\nu+1}+\la\omega_{\nu+1},I\ra
+\sum_{i\in \N} \la A_{i}^{\nu+1}z_i, z_i\ra+P_{\nu+1},
\label{4.2}
\end{equation}
 satisfying
\begin{equation}\label{3.6}
\max_{l\le \bar d^2}|\frac{\partial^l(\omega_{\nu+1}(\xi)-\omega_\nu(\xi))}
{\partial\xi^l}|\le \epsilon_{\nu}, \;\;
{\rm max}_{|l|\le \bar d^2}|\frac{\partial^{l}(A_{i}^{\nu+1}(\xi)-
A_{i}^\nu)}{\partial\xi^{l}}| \le \epsilon_{\nu} i^{-\delta},
\end{equation}

\begin{equation}
\|X_{P_{\nu+1}}\|_{D_{\nu+1},\Cal O_{\nu+1}}^{\bar a,\rho}
\le \epsilon_{\nu+1}.
\label{4.3}\end{equation}
\end{Lemma}

\subsection{ Convergence}

Suppose that the assumptions of Theorem 1 are satisfied. To apply the iteration lemma with $\nu=0$, recall that 
$$\epsilon_0=\epsilon, \gamma_0=\gamma, s_0=s,L_0=L,  N_0=N, P_0=P,$$

\begin{displaymath}
\Cal O_0=  
\left\{\xi\in \Cal O: 
|\; \stackrel{\scriptstyle |\la k,\omega\ra^{-1}|<\frac 
{|k|^{\tau}}{\gamma },
 \|(\la k,\omega\ra I_{d_i}+A_iJ_{d_i} )^{-1}\| < (\frac 
{|k|^\tau}\gamma )^{\bar d},\, \mbox{or}}
 {\scriptstyle \|(\la k,\omega\ra I_{d_id_j}+(A_iJ_{d_i})\otimes I_{d_j}
-I_{d_i}\otimes
(J_{d_j}A_j ))^{-1}\| <  (\frac {|k|^{\tau}}
\gamma)^{\bar d^2}}\;\right\}, 
\end{displaymath}
(with  $\epsilon$ and $\gamma$ small enough). 
Inductively, we obtain
the following sequences
$$\Cal O_{\nu+1}\subset  \Cal O_{\nu},$$
$$\Psi^\nu=\Phi_1\circ\cdots\circ\Phi_\nu: 
 D_{\nu+1}\times \Cal O_{\nu} \to D_0, \nu\ge 0,$$
$$H\circ\Psi^\nu=H_{\nu+1}=N_{\nu+1}+P_{\nu+1}.$$
 
\sss
Let $\Cal O_{\gamma}=\cap_{\nu=0}^\infty\Cal O_\nu$.
As in \cite{P2}, thanks to Lemma \ref{(3.5)}, 
we may conclude that  $N_\nu, \Psi^\nu, D\Psi^\nu, \omega_{\nu+1}$ converge uniformly
on $D_\infty\times \Cal O_\gamma=D(\frac 12 r, 0,0)\times
\Cal O_\gamma$ with
$$N_\infty=e_\infty+\la \omega_\infty,I\ra+\la A_\infty z,z\ra
=e_\infty+\la \omega_\infty,I\ra
+\sum_{i\in \N}\la A_i^\infty z_i, z_i\ra
,$$ 
Since 
$$\epsilon_{\nu+1} = 
c\gamma_{\nu}^{\gammab}\Gamma(r_{\nu}-r_{\nu+1})
\epsilon_{\nu}\le
(c\gamma^{\gammab}\Psi(r)\epsilon)^{(\frac 43)^\nu}.$$
It follows that $ \epsilon_{\nu+1}\to 0$  provided
 $\epsilon$ is sufficiently small.

\sss
Let $\phi^t_H$ be the flow of $X_H$. Since 
$H\circ\Psi^\nu=H_{\nu+1}$, we have that
\begin{equation}\phi^t_H
\circ \Psi^\nu= \Psi^\nu \circ \phi^t_{H_{\nu+1}}.
\label{4.10}
\end{equation}
The convergence of $\Psi^\nu,D\Psi^\nu,\omega_{\nu+1}, X_{H_{\nu+1}}$ implies that
 one can take limit in (\ref{4.10})  so as to get
\begin{equation}\label{4.11}
\phi^t_H\circ \Psi^\infty= \Psi^\infty\circ\phi^t_{H_\infty},
\end{equation}
on $D(\frac 12r, 0,0)\times \Cal O_\gamma$, with
$$
\Psi^\infty:D(\frac 12r,0,0)\times\Cal O_\gamma
\to   \Cal P_{a, \rho}\times \R^d.
$$

\sss
From (\ref{4.11}) it follows that 
$$\phi^t_H(\Psi^\infty(\T^d\times \{
\xi\}))=\Psi^\infty\phi^t_{N_\infty}
(\T^d\times \{\xi\})=\Psi^\infty (\T^d\times \{\xi\}),$$
for $\xi\in \Cal O_\gamma$.
This means that 
$\Psi^{\infty} (\T^d\times \{\omega\})$ is an embedded torus invariant 
for the original perturbed Hamiltonian
system at $\xi\in \Cal O_\gamma.$
We  remark here the frequencies $\omega_\infty(\xi)$
associated to $\Psi^{\infty} (\T^d\times\{\xi\})$
 is  slightly different from $\xi$. 
The normal behaviour of the invariant torus 
is governed by the matrix $A^{\infty}_i=\sum_{\nu\in \N}A^\nu_i$.    
\qed

\section{ Measure Estimates}\label{measureestimate}

At each KAM step, we have to exclude the following resonant 
set of $\xi$'s: $$\Cal R^{\nu}=\bigcup_{|k|>K_{\nu},i,j}(\Cal R_k^{\nu} \cup\Cal
R_{ki}^{\nu}
\cup\Cal R_{kij}^{\nu})\ ,$$ the sets $ \Cal R_k^{\nu}, \Cal R_{ki}^{\nu},
\Cal R_{kij}^{\nu}$ being respectively
\begin{eqnarray}
& &\{\xi\in \Cal O_\nu:  |\la k,\omega_{\nu}\ra^{-1}|>\frac 
{|k|^{\tau}}{\gamma_{\nu}}\},\quad
 \{\xi\in \Cal O_\nu:\,\, \|\Cal M_1^{-1}\|> (\frac 
{|k|^{\tau}}{\gamma_{\nu}})^{\bar d}\},\nonumber\\
& & \mbox{and}\;\; \{\omega\in \Cal O_\nu:\,\, \|\Cal M_2^{-1}\|> (\frac
{|k|^{\tau}}{\gamma_{\nu}})^{\bar d^2}\},
\end{eqnarray}
where
\begin{eqnarray} 
&\Cal M_1 &= \la k,\omega_{\nu}\ra I_{d_i}+A_i^{\nu}J_{d_i}\nonumber\\
&\Cal M_2 & = \la k,\omega_{\nu}\ra I_{d_id_j}+
(A_j^{\nu}J_{d_j})\otimes I_{d_i}-I_{d_j}\otimes
(J_{d_i}A_i^{\nu} )
.\label{kij}
\end{eqnarray}
We include in the 
set $\{\xi\in \Cal O: \|M(\omega)^{-1}\|>C\} $ 
also the $\xi$'s for which $M$ is not invertible.
Remind that $\omega_{\nu}(\xi)=\xi+
\sum_{j=0}^{\nu-1} P_{000}^j(\xi)$ with\footnote{Recall (\ref{Rest}), 
(\ref{3.6}).
 }
$|\sum P_{000}^j(\xi)|_{C^{\bar d^2}}\le \epsilon$,
$A_i^{\nu}= A_i+2\sum_\nu R^{0,\nu}_{ii}$ with $\|\sum_\nu R^{0,\nu}_{ii}\|
=O(\epsilon i^{-\delta})$.

\begin{Lemma}\label{Rkij} There is a constant $K_0$ such that,
for any $i,j,$ and $ |k|>K_0$,
\[{\rm meas }(\Cal R_{k}^{\nu}\cup\Cal R_{ki}^{\nu}
 \cup\Cal R_{kij}^{\nu})\lep 
\frac{\gamma}{ |k|^{\tau-1}}.\]
\end{Lemma}
\Proof
 As it is well known  
\[{\rm meas}\,(\Cal R_{k}^{\nu})\le \frac{\gamma_{\nu}}{|k|^\tau}.\]
The set $\Cal R_{ki}^{\nu}$ is empty
if $i>{\rm const}\;|k|$,  while, if $i\le {\rm const}\;|k|,$
from Lemmata \ref{3}, \ref{A.1} there follows that
\[{\rm meas }(\Cal R_{ki}^{\nu})\lep
\frac{\gamma_{\nu}}{ |k|^{\tau-1}}.\]

\sss
We now give a detailed proof for the most complicated estimate, i.e., the estimate
on the measure of the set
$\Cal R_{kij}^{\nu}$. Rewrite $\Cal M_2$ as 
\[\Cal M_2
\equiv \Cal A_{ij}+\Cal B_{ij}^{\nu},
\]  with
\begin{equation}\label{calaij}
\Cal A_{ij}= \la k,\omega_{\nu+1}\ra I_{d_id_j}+\lambda_j\,{\rm Diag}
(I_{d_j/2}, -I_{d_j/2})\otimes I_{d_i}
-\lambda_i I_{d_j}\otimes
{\rm Diag}\;(-I_{d_i/2}, I_{d_i/2}).
\end{equation}
The matrix $\Cal A_{ij}$ is diagonal with entries 
$\lambda_{kij}=\la k,\omega_{\nu}\ra\pm \lambda_i\pm \lambda_j$ 
in the diagonal where $\lambda_i, \lambda_j$ are given in (\ref{asymp1}) and
$\pm$ sign depends on the position.
 $\Cal B_{ij}^{\nu}$ is a matrix of  size $O(i^{-\delta}+
 j^{-\delta})$ since $ A_i^{\nu}=
A_i+ B_i+O(i^{-\delta})=A_i+O(i^{-\delta}) $ by (\ref{asymp2}) 
and (\ref{Rest}).

\ms
In the rest of the proof we drop in the notation the indices $i,j$ since they
 are fixed.
Now either all $\lambda_{kij}\le|k|$ or there are some diagonal elements 
$\lambda_{kij}> |k|$. We first consider the latter case.
 By permuting rows and columns, we can find  two non-singular
matrices $Q_1, Q_2$ with elements 1 or 0 such that
\begin{equation}\label{7.3}
Q_1(\Cal A +\Cal B^{\nu})Q_2
 =\left(\begin{array}{cc}
A_{11} & 0\\
0 & A_{22}
\end{array}\right)+
 \left(\begin{array}{cc}
\tilde B_{11} & \tilde B_{12}\\
\tilde B_{21} & \tilde B_{22}
\end{array}\right)
\end{equation}
where $A_{11}, A_{22}$ are diagonal matrices and $A_{11}$
 contains all diagonal elements $\lambda_{kij}$ which are bigger than
$|k|$. Moreover, defining $Q_3, Q_4, D$ as
\[Q_3=\left(\begin{array}{cc}
I & \tilde 0\\
 -\tilde B_{21}( A_{11}+\tilde B_{11})^{-1}
& I
\end{array}\right)
,\quad
Q_4=\left(\begin{array}{cc}
I & -( A_{11}+\tilde B_{11})^{-1}\tilde B_{12}\\
0 & I
\end{array} \right ),\]and
\begin{equation}
\label{D}
D=A_{22}+ \tilde B_{22}-\tilde B_{21}(A_{11}+\tilde B_{11})^{-1}
\tilde B_{12}= A_{22}+O(i^{-\delta}+j^{-\delta}),
\end{equation}
we have
\begin{equation} Q_3Q_1\left(\Cal A +\Cal B^{\nu+1}\right)Q_2Q_4=
\left(\begin{array}{cc}
A_{11}+B_{11} & 0\\
0 & D
\end{array}\right)
\end{equation}
For $\xi\in \Cal O$ such that $D$ is invertible,
we have 
\begin{equation}\label{inverse}
(\Cal A+\Cal B^{\nu})^{-1}=Q_2Q_4\left(\begin{array}{cc}
( A_{11}+B_{11})^{-1} & 0\\
0 & D^{-1}
\end{array}\right)Q_3Q_1.
\end{equation}
Since the norm of $Q_1, Q_2, Q_3, Q_4,  (A_{11}+B_{11})^{-1}$ are uniformly
bounded,
 it follows from (\ref{inverse}) that
\begin{equation}\label{7.6}
\{\xi\in \Cal O_\nu: 
\|(\Cal A+\Cal B^{\nu})^{-1}\|
> (\frac
{|k|^{\tau}}{\gamma_{\nu}})^{\bar d^2}\}\nonumber\\
 \subset
\{\xi\in \Cal O_{\nu}:
\|D^{-1}\| \gep (\frac
{|k|^{\tau}}{\gamma_{\nu}})^{\bar d^2}\}.
\end{equation}
If all $\lambda_{kij}\lep |k|$ we simply take $D=\Cal A+{\Cal B}^{\nu}$.
Since all elements in 
$D$ are of size $O(|k|)$, by Lemma \ref{3} in the 
Appendix, we have 
\begin{equation}\label{7.6'}
\{\xi\in \Cal O_{\nu}:
\|D^{-1}\|\gep \;( \frac
{|k|^{\tau}}{\gamma_{\nu}})^{\bar d^2}\}
\subset 
\{\xi\in \Cal O_{\nu}:
|\det D|\lep    (\frac{\gamma_{\nu}}
{|k|^{\tau-1}})^{\bar d^2}\}.
\end{equation}

\sss
Let $N$ denote the dimension of  $D$ (which is not bigger than
 $ \bar d^2$).
Since $D=A_{22}+O(i^{-\delta}+j^{-\delta})$, 
the $N^{\rm th}$ order derivative of $\det D$ with respective to some 
$\xi_i$ is bounded away from zero by
$\frac 1{2d} |k|^{N}$ (provided $|k|$ is bigger enough).
From (\ref{7.6}), (\ref{7.6'}) and  Lemma \ref{A.1}, it  follows  that
\begin{eqnarray}\label{7.6''}
{\rm meas}\,\Cal R_{kij}^{\nu}&=&
{\rm meas} \,\{\xi\in \Cal O_\nu: 
\|(\Cal A+\Cal B^{\nu})^{-1}\|
> (\frac
{|k|^{\tau}}{\gamma_{\nu}})^{\bar d^2}\}
\nonumber\\
&\le & {\rm meas}\{\xi\in \Cal O_{\nu}:
|\det D|\lep    (\frac{\gamma_{\nu}}
{|k|^{\tau-1}})^{\bar d^2}\}\nonumber\\
&\lep& (\frac{\gamma_{\nu}}{|k|^{\tau-1}})^{\frac {\bar d^2}N}
\lep \frac{\gamma}{|k|^{\tau-1}}.
\end{eqnarray}
This proofs the lemma.
\qed

\begin{Lemma}\label{empty}
If $i\gep |k|$, then $\Cal R^{\nu}_{ki}=\emptyset$;
If $\max\{i, j\}\gep |k|^{\frac 1{b-1}}, i\ne j$ for $b>1$ or 
$|i-j|>{\rm const}\; |k|$ for $b=1$, then $\Cal R^{\nu}_{kij}=\emptyset$
where the constant {\rm c} depends on the diameter of $\Cal O$.
\end{Lemma} 
\Proof As above, we only consider  the most complicated case, i.e., the case of $\Cal
R^{\nu}_{kij}$. Notice that  $\max\{i, j\}> {\rm const}\;|k|^{\frac 1{b-1}}$ for
$b>1$ or 
$|i-j|>{\rm const}\; |k|$ for $b=1$ implies 
\begin{eqnarray}
|\lambda_i\pm \lambda_j|&=&
(j^b-i^b)(1+O(i^{-\delta}+j^{-\delta}))
\nonumber\\
&\ge & \frac 12 |j-i|(i^{b-1}+j^{b-1})(1+O(i^{-\delta}+j^{-\delta}))
\ge {\rm const}\;|k|.\end{eqnarray}
It follows that $\Cal A_{ij}$ defined in (\ref{calaij}) is invertible and
\[\|( \Cal A_{ij})^{-1}\|<|k|^{-1}.\]
By Neumann series, 
we have $\|(\Cal A_{ij}+\Cal B_{ij}^{\nu})^{-1}\|< 2|k|^{-1}$ for large $k$ (say 
$|k|>K_0$),
i.e,  $\Cal R^{\nu}_{kij}=\emptyset$.
\qed

\begin{Lemma}
For $b\ge 1$, we have
 \[{\rm meas}(\bigcup_{\nu\ge 0}\Cal R^{\nu})
={\rm meas}\;\bigcup_{\nu, |k|>K_{\nu},i,j}(\Cal R_k^{\nu} \cup\Cal R_{ki}^{\nu}
\cup\Cal R_{kij}^{\nu})\lep \gamma^{\frac 
\delta{1+\delta}}. \]
\end{Lemma}
\Proof The measure estimates for $\Cal R^0$
comes from our assumption (\ref{ndc}). We then consider the estimate
\[{\rm meas}(\bigcup_\nu\bigcup_{|k|>K_{\nu}}\bigcup_{i,j}\Cal R^{\nu}_{kij}) \ ,\]
which is the most complicate one.

\sss
Let us consider separately the case $b>1$ and the case $b=1$.
We first consider $b>1$.
By Lemmata \ref{Rkij}, \ref{empty}, 
if  $|k|>K_0$ and $i\ne j$, we have
\begin{equation}\label{inej}
{\rm meas}\;(\bigcup_{i\ne j}\Cal R_{kij})=
{\rm meas}\;(\bigcup_{i\ne j; i,j<C|k|^{\frac 1{b-1}}}\Cal R^k_{ij})
\lep \frac{|k|^{\frac 2{b-1}}
\gamma }{|k|^{ \tau-1}}\gep
\frac\gamma {|k|^{ \tau-1-\frac 2{b-1}}}.  \end{equation}
For $i=j$.
As in Lemma \ref{Rkij}, we can find $Q_1, Q_2$ so that (\ref{7.3}) holds with 
 the diagonal elements of $A_{11}$ being
$<k, \omega_{\nu}>\pm 2\lambda_i$ and $A_{22}= <k, \omega_{\nu}>I$.
Repeating the arguments in Lemma \ref{Rkij}, we get (\ref{7.6'}) and   
\begin{eqnarray}\label{7.13}
 \Cal R_{kii}^{\nu} &\subset&
\{\xi:
|\det D|\lep
(\frac{\gamma_{\nu}}
{|k|^{\tau-1}})^{\bar d^2}\}\nonumber\\
&=&
\{\xi:\prod |\la k,\omega_{\nu}\ra+O(i^{-\delta})|
\lep
(\frac{\gamma_{\nu}}
{|k|^{\tau-1}})^{\bar d^2}\}\nonumber\\
&\subset&
\{\xi: |\la k, \omega_{\nu}\ra |\lep( \frac \gamma{|k|^{\tau-1}}
+\frac 1{i^{\delta}})\}\equiv \Cal Q_{ii}^k.\end{eqnarray}
Since  $\Cal Q_{ii}^k\subset \Cal Q_{i_0i_0}^k $ for $i\ge i_0$, using 
(\ref{7.6''}), we find
that
\[ {\rm meas}\;(\bigcup_i \Cal R_{kii})
 \le \sum_{i<i_0}|\Cal R_{kii} |+|\Cal Q_{i_0i_0}^k|
\lep (\frac{i_0\gamma} {|k|^{\tau-1}}+\frac 1{i_0^{-\delta}})\]
for any $i_0$.
Following P\"oschel (\cite{P2}), we choose
 $i_0= (\frac {|k|^{\tau-1}}{\gamma})^\frac 1{1+\delta}$, 
so that
\begin{equation}\label{7.13'}
 {\rm meas }\;(\bigcup_i \Cal R_{kii}| \lep 
(\frac {\gamma}{|k|^{\tau-1}})^ {\frac\delta{1+\delta}}\end{equation}
Let  $\tau>\max\{d+2+\frac 2{b-1},
(d+1)^{\frac{1+\delta}\delta}+1\}$. As in (\ref{inej}), (\ref{7.13'}), we find

\begin{eqnarray}
& &
{\rm meas}\;(\bigcup_{|k|>K_\nu}\bigcup_{i, j}\Cal R_{kij}^{\nu}
(\gamma_{\nu}))
={\rm meas}\;(\bigcup_{|k|>K_\nu}\bigcup_{i\ne j}\Cal R_{kij}^{\nu}
(\gamma_{\nu}))\nonumber\\
\ \ \ & & +{\rm meas}\;(\cup_{|k|>K_\nu}\bigcup_{i }\Cal R_{kii}^{\nu}(\gamma_{\nu+1}))
\lep K_\nu^{-1}\gamma^{\frac\delta{1+\delta}}.\nonumber
\end{eqnarray}
The quantity
${\rm meas}(\bigcup_\nu\bigcup_{|k|>K_{\nu}}\bigcup_{i,j}
\Cal R^{\nu}_{kij})$ is then
bounded by

\begin{equation}
\sum_{\nu\ge 1} {\rm meas}(
\bigcup_{|k|>K_\nu}\bigcup_{i, j}\Cal R_{kij}^{\nu}(\gamma_{\nu}))
\lep \gamma^{\frac \delta{1+\delta}}\sum_{\nu\ge 0}K_{\nu}^{-1}
\lep \gamma^{\frac \delta{1+\delta}},
\end{equation}
provided $\tau>\max\{d+2+\frac 2{b-1},
(d+1)^{\frac{1+\delta}\delta}+1\}$. This concludes the proof for $b>1$.

\ms
Consider now $b=1$. Without loss of generality, we assume $j\ge i$ and
 $j=i+m$. Note that  Lemma 
{\ref{empty} implies $\Cal R^k_{ij}=\emptyset$ for $m>C|k|$. 
Following the scheme of the above proof, we 
find
\begin{eqnarray}
\bigcup_{k,i,j}\Cal R_{kij}&=&\bigcup_{k,i,m}\Cal R_{ki,i+m}=
\bigcup_{k, m<C|k|}\bigcup_i\Cal R_{ki,i+m}\nonumber\\
&\subset &\bigcup_{k, m<C|k|}(\bigcup_{i<i_0}\Cal R_{ki_0,i_0+m}\cup 
\Cal Q_{ki_0,i_0+m} ).
\end{eqnarray}
where 
\[\Cal Q_{k i_0, i_0+m}=\{\xi: |\la k, \omega_{\nu}\ra +m
|\lep( \frac \gamma{|k|^{\tau-1}}
+\frac 1{i^{-\delta}})\}.
\]
Again, taking $i_0^{1+\delta}= \frac{|k|^{\tau-1}}{\gamma},$
we have, for fixed $k$,
\begin{eqnarray}
|\bigcup_{i,j}\Cal R_{ij}^k| &\lep& \sum_{m<C|k|}(\frac{i_0\gamma}{|k|^{\tau-1}}
+i_0^{-\delta})\nonumber\\
&\lep& |k|( \frac \gamma {|k|^{\tau-1}})^{\frac \delta{1+\delta}}\ .
\end{eqnarray}
As in the case $b>1$, we have that  
${\rm meas}(\bigcup_\nu\bigcup_{|k|>K_{\nu}}\bigcup_{i,j}\Cal R^{\nu}_{kij})$
is bounded by
$O(\gamma^{\frac\delta {1+\delta}})$
if $\tau> (d+1)^{\frac {1+\delta}\delta} +1$.
\qed

\noindent
{\bf Remark} \  In (\ref{7.13}), $|\det D|=\prod|\ko +O(i^{-\delta})|$
(guaranteed by the regularity property) is crucial for the proof. But it is
not necessary for the periodic solution case, i.e., $d=1$.
since  $\Cal R_{k, i+m, i}^{\nu}
=\emptyset$ if, $ i\gep> |k|\ll 1$ are sufficiently large.

\section{Appendix}

\sss
{\it Proof of Proposition \ref{prop}} From the hypotheses there follows that the
eigenfuctions $\phi_n$ are analytic (respectively, smooth) and bounded with, in
particular, 
$$\sup_{\Bbb R} (|\phi_n'|+|\phi_n''|)\le\  {\rm const}\  \mu_n\ .$$
Thus, the sum defining $u(t,x)$ is uniformly convergent in $I\times [0,2\pi]$. Since
$$
\frac{\partial G}{\partial q_n} = - \frac{1}{\sqrt{\lambda_n}} \int f(\sum_k
\frac{q_k}{\sqrt{\lambda_k}} \phi_k) \phi_n\ ,
$$
one has 
$$
|q_n|\le {\rm const}\ \frac{e^{-n\rho}}{n^a}\ ,\quad
|\dot q_n|\le {\rm const}\ \lambda_n \frac{e^{-n\rho}}{n^a}
\le {\rm const}\ \frac{e^{-n\rho}}{n^{a-1}}\ ,
$$
$$|\ddot q_n|\le {\rm const}\ \frac{e^{-n\rho}}{n^{a+1}}\ .
$$
Thus (if $a$ is big enough, in the smooth case) $u(t,x)$ is a $C^2$ function and
\begin{eqnarray}
u_{tt}+Au &=& \sum \frac{\ddot q_n}{\sqrt{\lambda}} \phi_n +
\frac{q_n}{\sqrt{\lambda_n}} A \phi_n \nonumber\\
&=& \sum \Big( \int f(u) \phi_n\Big) \phi_n = f(u)\ ,\nonumber\\
\end{eqnarray}
where in the last equality we used the fact that $f(u)$ is a smooth periodic
function. 
\qed

\begin{Lemma}
\[\|FG\|_{D(r,s)}\le  \|F\|_{D(r,s)} \|G\|_{D(r,s)}.\]
\end{Lemma}
\Proof 
Since $(FG)_{klp}=\sum_{l}F_{k-k',l-l',p-p'}G_{k'l'p'}$. 
we have that
\begin{eqnarray}
\|FG\|_{D(r,s)}&=&\sup_D \sum_{klp}|(FG)_{klp}| \ |y|^l\ |z^\alpha|
e^{|k|r}
\nonumber\\
&\le &\sup_D\sum_{klp} \sum_{l'}|F_{k-k',l-l',p-p'}G_{k'l'p'}|\  
|y|^l|z^\alpha| e^{|k|r}\nonumber\\
&=& \|F\|_{D(r,s)} \|G\|_{D_(r,s)}
\end{eqnarray}
and the proof is finished.
\qed

\begin{Lemma}(Cauchy inequalities)
\[\|F_{\theta_i}\|_{D(r-\sigma, s)}
\le c\sigma^{-1}\|F\|_{D(r,s)},\]
and
\[ \|F_{I}\|_{D(r,\frac 12s)}\le 
2\frac 1{s^2} \|F\|_{D(r, s)},\ \ \  
\|F_{z_n}\|_{D(r,\frac 12s)}\le 
2\frac {n^ae^{n\rho}}{s} \|F\|_{D(r,s)}
\]
\end{Lemma}
Let $\{\cdot, \cdot\}$ is Poisson bracket of smooth functions
\begin{equation}
\{F, G\}=
\sum (\frac {\partial F}{\partial \theta_i}
\frac {\partial G}{\partial I_i}-
\frac {\partial F}{\partial I_i}
\frac {\partial G}{\partial \theta_i})+\sum_{i\in \N}
 \la\frac{\partial F}{\partial z_i}, {\rm i} J_{d_i}\frac{\partial G}{\partial z_i}
\ra,
\end{equation}
where $J_{d_i}$ are standard symplectic matrix in ${\Bbb R}^{d_i}$.

\begin{Lemma}\label{FG}
If 
\[ \|X_F\|_{r,s}<\epsilon',  
\|X_G\|_{r,s}<\epsilon'',\]
then
\[\|X_{\{F, G\}}\|_{ r-\sigma,\eta s}
\lep\sigma^{-1} \eta^{-2}\epsilon'\epsilon'', \quad \eta\ll 1.\]
\end{Lemma}
\Proof
Note  that  
\begin{eqnarray} \frac d{dz_n}\{F,G\}& &=\la F_{\theta z_n}, G_I\ra
+\la F_\theta,
 G_{Iz_n}\ra
-\la F_{Iz_n}, G_\theta\ra -\la F_I, G_{\theta z_n}\ra\nonumber\\
& & +\sum_{i\in \N}(\la F_{z_iz_n}, J_{d_i}G_{z_i}\ra+
\la F_{z_i}, J_{d_i}G_{z_iz_n}\ra )
\end{eqnarray}
Since
\begin{eqnarray}
\|\la F_{\theta z_n}, G_I\ra\|_{D(r-\sigma,s)}&\lep &
\sigma^{-1} \|F_{z_n}|\|\cdot \|G_y\|
\nonumber\\
\|\la F_\theta, G_{Iz_n}\ra \|_{D(r-\sigma, \frac 12 s)}&\lep &
s^{-2} \|F_{\theta}\|\cdot \|G_{z_n}\|
\nonumber\\
\|\la F_{Iz_n}, G_\theta\ra \|_{D(r, \frac 12 s)}&\lep &
s^{-2} \|F_{z_n}\|\cdot \|G_\theta\|
\nonumber\\
\|\la F_I, G_{\theta z_n}\ra \|_{D(r-\sigma, s)}&\lep &
\sigma^{-1} \|F_{I}\|\cdot \|G_{z_n}\|
\nonumber\\
\|\la F_{z_iz_n}, J_{d_i}G_{z_i}\ra \|_{D(r, \frac 12s)}&\lep &
s^{-1} \|F_{z_n}\|\cdot \|G_{z_i}\| i^ae^{i\rho}
\nonumber\\
\|\la F_{z_iz_n}, J_{d_i}G_{z_i}\ra \|_{D(r, \frac 12s)}&\lep &
s^{-1} \|F_{z_n}\|\cdot \|G_{z_i}\|i^ae^{i\rho}
\end{eqnarray}
it follows from the definition of the weighted norm(see (\ref{weightednorm})),
that 
\[\|X_{\{F, G\}}\|_{ r-\sigma, \eta s}
\lep \sigma^{-1}\eta^{-2}\epsilon'\epsilon''.\]
In particular, if $\eta\sim \epsilon^{\frac 13}, \epsilon', \epsilon''\sim 
\epsilon$, we have $\|X_{\{F, G\}}\|_{ r-\sigma, \eta s}\sim 
\epsilon^{\frac 43}$.
\qed

\begin{Lemma}\label{extension1}
Let $\Cal O$ be a compact set  in ${\Bbb R}^d$ for which (\ref{ODC}) holds. 
Suppose that
$f(\xi)$ and $\omega(\xi)$
are  $C^m$ Whitney-smooth function in $\xi\in \Cal O$  with $C_W^m$ norm
 bounded by $L$.
Then 
\[g( \xi)\equiv\frac{f(\xi)}{\la k, \omega(\xi)\ra}\]
is $C^m$ Whitney-smooth in  $\Cal O$
with\footnote{Recall the definition in (\ref{Falpha}).} 
\[\|g\|_{\Cal O}\lep \gammaa\kk L,\]
\end{Lemma}
\Proof
The proof follows directly from the definition of the Whitney's 
differentiability.
\qed
A Similar lemma for matrices holds:

\begin{Lemma}\label{extension2}
Let $\Cal O$ be a compact set in ${\Bbb R}^d$ for which (\ref{ODC}) holds. 
Suppose that
$B(\xi), A_i(\xi)$
are  $C^m$ Whitney-smooth matrices and $\omega(\xi)$ is 
a Withney-smooth function  in $\xi\in \Cal O$  bounded by $L$.
Then 
\[C(\xi)=B M^{-1},\]
is $C^m$ Whitney-smooth 
with 
\[\|F\|_{\Cal O}\lep \gammaa\kk L,\]
where $M$ stands for either
$\la k, \omega\ra  I_{d_i}+ A_iJ_{d_i}$ if
$B$ is $(d_i\times d_i)$-matrix, or
$\la k, \omega\ra  I_{d_id_j}+(A_iJ_{d_i})
\otimes I_{d_j}-I_{d_i}\otimes (J_{d_j}A_j)$
if $B$ is $(d_id_j\times d_id_j)$-matrix,
\end{Lemma}
For a  $N\times N$ matrix  $ M=(a_{ij})$,
 we denote by $| M|$  its 
determinant.  Consider $M$ as a linear operator on 
$(R^N, |\cdot|)$ where $| x|=\sum |x_i|$. Let 
$\|M\|$ be its operator norm.
It is known $\| M \|$ is equivalent to norm  $ \max |a_{ij}|$.
Since a constant depends only on the space dimension and two fixed
norms is irrelevant, 
we will simply denote  $\|M\|= \max |a_{ij}|$.

\begin{Lemma}\label{3}
Let $M$ be a $N\times N$ non-singular matrix with
 $\|M\|\lep |k|$, then
\[\{\omega: \|M^{-1}\|>h\}
\subset\{\omega:  |\det M|\lep \frac{|k|^{N-1}}h\}
\]\end{Lemma}
\Proof
 Firstly,  we note that if $M$ is a nonsingular $N\times N$ matrix with 
elements bounded by $|m_{ij}|\le m$, by Cramer rule,  the inverse of 
$M$ is $M^{-1}= \frac 1{|M|}{\rm adj}M$. Thus 
\[\|M^{-1}\|\lep \frac {m^{N-1}}{|{\rm det} M|}  \]
where the constant depends on $N$.
In particular, if $m={\rm const} |k|,  |{\rm Det} M|> 
 \frac{|k|^{N-1}}h $ then
\[ \|M^{-1}\|\lep h.\]
This proofs the lemma.
\qed
In order to estimate the measure of $\Cal R^{\nu+1}$, we need the
following lemma, which has been proven in \cite{XYQ} \cite{Y2}.
A similar estimate is also used by Bourgain \cite{B2}.

\begin{Lemma}\label{A.1}  
 Suppose that $g(u)$ is a $C^m$ function on  the closure
 $\bar I$, where \ $I\subset R^1$\  is  a finite interval.
Let\  $I_h=\{u |\ \ \ |g(u)|<h  \},\ \ \ h>0 .$\ If  for 
some constant $d >0$,
 $|g^{(m)}(u)| \ge d  $ for  all $u \in I$, then 
$ {\rm meas}\;(I_h) \le ch^{\frac1{m}}$
where 
 $ c=2(2+3+\cdots +m+d^{-1}).$
\end{Lemma}
For the proof of Lemma \ref{regularity}, we need the following
\begin{Lemma}\label{sum1}
\[ \sum_{j\in \Z}e^{-|n-j|r+\rho |j|}\le Ce^{\rho |n|}, 
\sum_{j, n\in \Z}|q_j|e^{-|n-j|r+ |n|\rho}\le C|q|_\rho\]
if  $\rho <r, q\in \Cal Z_{\rho}$ where $C$ depends on $r-\rho$.
\end{Lemma}

\begin{Lemma}\label{sum2}
\[ \sum_{j\in \Z}(1+|n-j|)^{-K} |j|^a\lep |n|^a, 
\sum_{j, n\in \Z}|q_j|(1+|n-j|)^{-k} |n|^a\le C|q|_a\]
if  $K>a+1, q\in \Cal Z_{a,\rho=0}$ where $C$ depends on $K-a-1$.
\end{Lemma}

\ms 
The proofs of the above two lemmata are elementary and we omit them.

\bs
\noindent
{\it Proof of Lemma \ref{regularity}}:
 Here we give a direct proof.
It is clearly enough to consider the case of $f(u)$ being a monomial $u^{N+1}$ for
some  $N\ge 1$. 
From (\ref{tildeG}), one can see that the regularity of $G$ implies
the regularity of $\tilde G$. In the following, we shall give the proof 
for $G$.
\sss
Suppose that the potential $V(x)$ is analytic
in $|{\rm Im} x|<r$ (respectively, belongs to 
Sobolev space $H^K$)
then the eigenfunctions are analytic
in $|{\rm Im} x|<r$ (respectively, belong  to  $H^{K+2}$).  If we let 
$\phi_i(x)=\sum a_i^n\nx $ then(see, e.g.,\cite{CW})
\[|a_{i}^{n}|\lep e^{-|i-n|r} \quad {\rm respectively}
\quad |a_{i}^{n}|\lep(1+|n-i|
^{-K-2}).\] 

Recall that
\[G(q)=\sum_{i_0,\cdots ,i_N}C_{i_0\cdots i_N}\frac{q_{i_0}\cdots q_{i_N}}
{\sqrt{\lambda_{i_0}\cdots \lambda_{i_N}}}\]
where 
\[C_{i_0\cdots i_N}=
\int_{T^1}\phi_{i_0}\cdots\phi_{i_N}dx=\sum_{n_0+n_1+\cdots+n_N=0}
(\prod_{s=0}^Na_{i_s}^{n_s}),
\]
with
$|a_{i_s}^{n_s}|\lep e^{-|i_s-n_s|r}$ (respectively,
$|a_{i_s}^{n_s}|\lep(1+|n_s-i_s|
^{-K-2})$.

In what follows, we assume either $a=0, \rho>0$ or $a>0, \rho=0$.
Since 
\[G_{q_j}=(N+1)\sum_{i_1,\cdots, i_N }
C_{ji_1\cdots i_N}
\frac{q_{i_1}\cdots q_{i_N}}
{\sqrt{\lambda_{j}\lambda_{i_1}\cdots \lambda_{i_N}}}
\]
it follows that
\begin{eqnarray} 
& &\|G_q\|_{a+\frac12, \rho}= \|G_{q_0}\|+\sum_{j\ge 1}
|G_{q_j}| |j|^{a+\frac 12}e^{j\rho}\nonumber\\
&\lep&\sum_{j,i_1,\cdots,i_N,\atop
 n_0+\cdots +n_N=0}|a_{j}^{n_0}|j^{a}e^{|j|\rho}
(\prod_{s=1}^N|a_{i_s}^{n_s}q_{i_s}|) 
\nonumber\\
&\lep &
\sum_{j,i_1, \cdots, i_N;\atop n_0+\cdots +n_N=0}(1+|j-n_0|)^{-N}|j|^a
e^{|j|\rho-|n_0-j|r}
\left(\prod_{s=1}^N(1+|n_s-i_s|)^{-K-2} 
e^{-|n_s-i_s|r}|q_{i_s}|\right)\nonumber\\
&\lep &
\sum_{i_1, \cdots, i_N;\atop n_0+\cdots +n_N=0}|n_0|^ae^{|n_0|\rho}
\left(\prod_{s=1}^N(1+|n_s-i_s|)^{-K-2}e^{-|n_s-i_s|r}|q_{i_s}|\right)\nonumber\\
&\lep &
\sum_{i_1, \cdots, i_N; \atop n_1, \cdots,n_N}(|\sum_{s=1}^Nn_s|)^a
e^{|\sum_{s=1}^N n_s|\rho}
\left(\prod_{s=1}^N(1+|n_s-i_s|)^{-K-2}e^{-|n_s-i_s|r}|q_{i_s}|\right)\nonumber\\
&\lep &
\sum_{i_1, \cdots, i_N;\atop
 n_1,\cdots,n_N}
\left(\prod_{s=1}^N(1+|n_s-i_s|)^{-K-2}  |n_s|^ae^{-|n_s-i_s|r+|n_s|\rho}|q_{i_s}|\right)\nonumber\\
&\lep &
\sum_{i_1, \cdots, i_N}
\left(\prod_{s=1}^N|i_s|^a e^{|i_s|\rho}|q_{i_s}|\right)\nonumber\\
&\lep &
\prod_{s=1}^N
\left(\sum_{i_s}|i_s|^ae^{|i_s|\rho}|q_{i_s}|\right)\lep |q|_{a,\rho}^N.
\end{eqnarray}
\qed

\end{document}